\documentclass[10pt]{article}
\usepackage{amssymb}
\usepackage{latexsym}
\usepackage{epsfig}
\usepackage{float}
\usepackage{graphicx}
\usepackage{epstopdf}
\usepackage{amsmath}
\usepackage{slashed}
\begin{document}
\vspace*{3cm}
\begin{center}
{\bf{\Large   Impact of the vector dark matter on polarization of the CMB photon }}
\vskip 4em{ {\bf  S. Modares Vamegh$^a$} \footnote{e-mail:s.modares@shirazu.ac.ir}\: ,
\: {\bf M. Haghighat$^{a,b}$}\footnote{e-mail:m.haghighat@shirazu.ac.ir}}\: ,
\: {\bf  S. Mahmoudi$^a$} \footnote{e-mail:s.mahmoudi@shirazu.ac.ir}\: and\: {\bf  R. Mohammadi$^{c,d}$} \footnote{e-mail:rmohammadi@ipm.ir} 
\vskip 1em
$^a$ Department of Physics, Shiraz University 71454, Shiraz, Iran\\
$^b$ ISC, 71946-94173, Shiraz, Iran\\

$^c$Iranian National Science and Technology Museum (INMOST),
 11369-14611, Tehran, Iran\\
$^d$School of Astronomy, Institute for Research in Fundamental Sciences (IPM),
19395-5531, Tehran, Iran
\end{center}
\vspace*{1.8cm}

\begin{abstract}
We consider a vector dark matter (VDM) with a direct coupling with photon.  We examine the effect of such an interaction on the CMB polarization to put new constrains on the properties of the DM particles.   We show that a partially polarized VDM of the order of temperature fluctuation with a quadrupole distribution leads to a valuable CP for the CMB.   In different DM-models the DM-masses range from few $eV$ to a few $TeV$.  We show that the CP  angular power spectrum depends on the mass of VDM as $C^{(S)}_{Vl}\propto 1/m_{_{V}}^6$ such that for  $m_{_{V}}=10eV-1keV$, the CP angular power spectrum is  $C^{(S)}_{Vl}\simeq 10^3- 10^{-11}{\rm nK^2}$.  Therefore, the light VDM with masses less than $10 eV$ leads to an unexpected very large CP which can be excluded from the acceptable range of the VDM masses.
\end{abstract}
\section{Introduction}
There are different theoretical and experimental researches on the dark matter (DM) which contributes  about $25 \% $ of the whole energy density of the Universe.
Although, many cosmological and astrophysical observations such as  velocity curves of spinning galaxies \cite{rotatecurve},  gravitational lensing \cite{gravitlens} and structure formation \cite{strucform} confirm the existence of DM in the universe but its nature is still unknown.  However,
Massive astrophysical compact halo object (MACHO) \cite{macho}, Gravitino \cite{gravitono}, Kaluza-Klein particle \cite{kaizu},  nuetrinos \cite{nuetrino}, weakly interacting massive particles  (WIMPs) \cite{WIMP1,WIMP2}, super-WIMPs \cite{SUPERWIMP1,SUPERWIMP2}, axions \cite{AXION} are considered as some of DM candidates.
Among these various candidates,  WIMPs and super-WIMPs  are particularly attractive candidates for DM at the electroweak scale.
The  WIMPs are in GeV mass scale and  fit to  the Cold DM paradigm.
They not only naturally describe the now-observed abundance of dark matter \cite{observaba} but also  hold all cosmological and astrophysical constraints imposed on dark matter experiments \cite{Wim}.
This is while, the super-WIMPs are known as warm DM candidate. They are lighter than the WIMPs and put in the range of $keV$ mass scale.
They can potentially explain the small-scale gravitational clustering properties and for above $\sim 3keV$ , these particles can easily describe the structure formation \cite{form}.
Detection of WIMPs  and super-WIMPs due to weakly interaction are naturally difficult nevertheless there are some suggested experiments for finding these particles such as XENON100 \cite{ XENON100I,XENON100II}, CDMS \cite{CDMS}, LUX \cite{LUX}, PAMELA \cite{PAMELAI,PAMELAII}, MAGIC \cite{MAGIC}, VERITAS \cite{VERITAS},  DAMA \cite{DAMA} and XMASS \cite{XMASS}.  Nontheless, even for very light dark matters below $keV$ there are models  for the Bosonic dark matters and also experimental direct detection for masses in $meV$ to $keV$ range  \cite{below keV}.
 On the other hand, Many experiments such as WMAP \cite{WMAP}, SPT \cite{SPT}, ACT \cite{ACT} and Planck \cite{Planck} are dealt with the CMB fluctuations and its polarization modes. In fact, the cosmological information is encoded in the temperature and polarization anisotropies of the CMB radiation. Meanwhile, at the recombination epoch according to the standard model of cosmology the Compton scatterings off particles such as electrons, protons, and so on  are the main source of the photons interaction.  However,  depending on the type of perturbations the Compton scattering can only produce the E and B-modes of linear polarization (LP) and not a radiation with CP. Furthermore,   temperature and the LP fluctuations of the CMB have been measured as $\frac{\delta T}{T_{\rm CMB}} \leq 10^{-5}$ while the current upper limit on the CP is about $\frac{\delta V}{T_{\rm CMB}}\leq 10^{-4}$ at the large scales \cite {tttv,tttv2,tttv3,tttv4}.
 Therefore, any new interaction beyond the standard cosmological model with  photons at the last scattering surface can provide a new source of information through producing different types of the CMB polarization.
For instance, in the Ref. \cite{lens1} it is shown that  the gravitational lensing by the matter distribution
can transform E-  into B-mode polarization of the CMB.
In \cite{nQED} the contribution of the non-commutative QED on the CMB Linear Polarization has been considered.
Non-linear photons interactions via Euler-Heisenberg effective Lagrangian is  introduced as a source of CP of CMB in Ref. \cite{photon2}.
In Refs. \cite{photonotrino,photonotrino2}  have been shown that the photon scattering from the cosmic neutrino background  can have a significant contribution on both LP and CP of the CMB.\\
However, a natural question can arise here, how much different DM-models can affect the CMB polarization.    In these models DM-candidates may be appear as fermions  or bosons with different spins \cite{scalar, vector, DMdipol1,DMdipol2,VDM2}.  For example, in  \cite{DMdipol1} a fermionic  DM has been considered as a dipolar DM which couples to photon through loops in the form of electric and magnetic dipole moments.  Based on this model,  the DM effects on the CMB photon polarization has been explored which is resulted in a  B-mode polarization in presence of primordial scalar perturbations with a bound on the mass and magnetic dipole moment of dipolar DM about $m\geq 50MeV$ and $M\leq10^{-16}e\,\,cm$, respectively \cite{DDMM}.

Here, we would like to focus on a model given in \cite{VDM2}  with a spin-1 bosonic vector-DM (VDM) which can directly couple to photon through a gauge invariant dimensionless coupling. As we know, the vector boson as the DM candidate has recently received much attention in various mass scales from a few eV to few TeV.
For instance,  in \cite{SVDM}  it is shown that the Higgs portal singlet VDM scenario with $60\,GeV\lesssim m_{_{V}}\lesssim 80\,GeV$ can naturally explain the $\gamma$-ray spectrum.
A vector WIMP with $100-1000\,GeV$ mass which acquires its mass from the gauge symmetry breaking at the TeV scale  has been considered in  \cite{mircle} .
A model for VDM with $m_{_{V}}>63GeV$ has been introduced in \cite{VDM3} where  the bound on the VDM mass has been obtained from the upper bound on the
invisible decay width of the Higgs boson.
In  \cite{VDM2040} it is shown that a VDM with $20-40\,GeV$ mass can provide an excellent description of the observed gamma-ray excess. In \cite{VDM2,VDM4} a model with two different VDM candidates  are considered in a $GeV$ mass scale to explain $3.5\,keV$ and $130\,GeV$ photon lines.
Meanwhile, there are many papers for the VDM in the low mass scale.  In  \cite{KVDM},  the constraints on the minimal model for VDM leads to $0.01-100\,keV$ mass range.  In \cite{SUPERWIMP1} the cosmological abundance, the background created by particle decays and the impact on
stellar processes due to cooling are analyzed  for the VDM with mass less than $100\,keV$.
The authors in \cite{XENON100II}  have obtained some bounds on the mass of VDM in range $8-125\,keV$.  However, a direct detection  with the XMASS-I liquid xenon detector for the mass of the VDM has been resulted in a  $40-120\,keV$ range \cite{XMASS} .

The paper is organized as follows.  Sec. II presents an overview of polarization and Boltzmann equations. Sec. III discusses the time evolution of Stockes parameters due to DM-photon interaction for forward scattering and damping terms. Sec IV details CP of CMB photons and  its power spectrum due to DM-photon scattering. Finally, we summarize our results and give conclusion in Sec. VI.
\section{BOLTZMANN EQUATIONS AND STOKES PARAMETERS}
 
 The polarization property is an important source to get information from photon as a quanta of electromagnetic field.
Of different formalism that can describe photon polarization, the Stokes parameters which describe partial polarization of photon in terms of 4 parameters, $I$, $Q$, $U$ and $V$ is proper for astronomical polarimetry where $I$ indicates the total intensity of radiation, $Q$ and $U$ show the linear polarization and $V$ denotes the difference between right- and left- handed CP.  However, an unpolarized photon characterizes with $Q=U=V=0$. Meanwhile, for an ensemble of photons the Stokes parameters can be described  as the components of photon density matrix $\rho$ as follows
 \begin{equation}\label{matrisro1}
\rho = \frac{1}{2}
\Bigg(
\begin{array}{rr}
I+Q \,\,& U-iV \\\\
U+iV\, & I-Q
\end{array}
\Bigg).
\end{equation}
Since the density matrix for a system of photons contains the same information as the four Stokes
 parameters one can examine the time evolution of the density matrix to find the polarization of the system.

For every component of the universe one can consider a statistical distribution function which depends on time, position and momentum  \cite{bol}.
The statistical equation that describes the features of macroscopic distribution of particles due to gravity and collisions is called Boltzmann equation which generally can be written as \cite{bol}
\begin{equation}
\frac{d f}{dt}= C[f],
\end{equation}
where the left-hand side deals with space-time properties, gravitational perturbutions around the homogeneous cosmology, is known as the Liouville term while the righ-hand side of this equation contains all possible collision terms. The time evolution of photon density matrix $\rho_{ij}$ can be considered as a generalized Boltzmann equation as follows \cite{stoke1}
\begin{equation}\label{roij}
 (2\pi)^3 \delta^3(0)(2k^0)\frac{d}{dt}\rho_{ij}({\bf{k}})=i\langle[H_I^0(t),D_{ij}^0({\bf{k}})]\rangle-\frac{1}{2}\int dt \langle[H_I^0(t),[H_I^0(0),D_{ij}^0({\bf{k}})]]\rangle,
\end{equation}
where $H_I^0 $ is the interacting Hamiltonian \cite{stoke1}, $D_{ij}^0({\bf{k}})=a_i^\dagger({\bf{k}})a_j({\bf{k}})$ is the number operator and $\langle\rangle$ indicates the expectation value. The first  term on the right-hand side of (\ref{roij}) is called forward scattering. The second one is known  as damping term which is related to higher order collisions. In the following section, we consider the interaction between VDM and the CMB photons to explore the possible effects of this interaction on the CMB  polarization.
\section{THE TIME EVOLUTION OF STOCKS PARAMETERS VIA  DM-PHOTON INTERACTION}
As we know the nature of dark matter is unknown. One possible candidate for explaining the unseen dark sector of the universe is spin-1 VDM.
Here, we use a model which is introduced in Ref. \cite{VDM2}. Comparing to the standard model, this model includes an extra neutral vector boson pair $V$ and $V'$  with masses $m_{_{V}}<m_{_{V'}}$. Due to $Z_{2}$ symmetry imposed on this model, vector boson $V$ is stable and can be a suitable candidate  for DM.  However,
 in this model  photon can directly couple to the  dark matter particles through a gauge invariant coupling as follows
\begin{eqnarray}\label{LI}
L_{I}=g_{_{V}}\cos \theta_{_{W}}F^{\mu\upsilon}V_{\mu}V'_{\nu}+g'_{_{V}}\cos \theta_{_{W}}\varepsilon^{\mu\nu\alpha\beta}F_{\mu\upsilon}V_{\alpha}V'_{\beta},
\end{eqnarray}
where $g_{_{V}}$ and $g'_{_{V}}$ are dimensionless coupling, $\theta_{_{W}} $ is the Weinberg angle, $V_{\mu}$  and $V'_{\mu}$ are VDMs and  $F_{\mu\nu}=\partial_{\mu}A_{\nu}-\partial_{\nu}A_{\mu}$ is the field strength of radiation.
In the rest of the paper, we explore the effects of VDM-photon scattering on the CMB polarization. To this end, we make use of (\ref{roij}) to calculate the time evolution of the component of the density matrix and show that how the interaction between photon and the VDM can affect the CMB polarization.
\begin{figure}
	\includegraphics[width=4in]{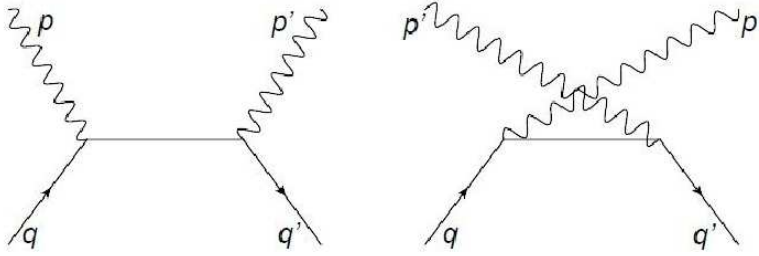}\\
	\caption{Typically Feynman diagrams for VDM-photon scattering}\label{diag1}
\end{figure}
\\
\subsection{FORWARD SCATTERING TERM}
As the largest term we first consider the  forward scattering one.  It can be shown that the time evolution of Stockes parameters due to the DM-photon forward scattering  Fig. \ref{diag1} with a completely unpolarized VDM and an arbitrary distribution function $\rho^{0}(\bf{x},\bf{q})$ leads to (see appendix A)
\begin{eqnarray}\label{unpol}
\dot{I}({\bf{k}})=0,\,\,\,\,\,\,\,\,\,\,\dot{V}({\bf{k}})=0,\nonumber\\
\dot{Q}({\bf{k}})=0,\,\,\,\,\,\,\,\,\,\dot{U}({\bf{k}})=0. \label{stoke12}
\end{eqnarray}
In the other words, (\ref{unpol}) means that an unpolarized VDM  can not generate any polarization for the CMB radiation.
However, the net polarization of the VDM depends on model in which it can be produced.  In fact, in different models there might be an imbalance in the production of each component in comparison with the other ones\cite{VDM-production}.  Here,  without referring to any specific model, we would like to assume that for some reason or fluctuation there is a small difference between the distribution densities of each mode of polarizations. To this end, we consider a polarized distribution for the  VDM  by taking  $\rho_{11}\neq\rho_{22}\neq\rho_{33}$ into account  where   $\rho_{11}$ and $\rho_{22}$ show the densities of  two transverse modes and  $\rho_{33}$  is  the density of the  longitudinal mode.  Meanwhile, one can define the density matrix elements for each components as  
\begin{equation}\label{matrisro1}
\rho_{rr'}(\bf{x},\bf{q}) =
\bigg\lbrace 
\begin{array}{rr}
{(\frac{1}{3}+\delta}_{r} )\,\,\rho^{0}({\bf{x}},{\bf{q}})\,& r=r' \\\\
0\,\,\,\,\,\,\,\,\,\,\,\,\,\,\,\,\,\,\,\, &  r\neq r' 
\end{array}
\end{equation}
where $|\delta_{r}|\ll 1$ shows a small deviation from the unpolarized case.  Now  the time evolution of the  Stocks parameters can be explored order by order in terms of $l$ the rank of the multipole expansion of the angular function in terms of the Legendre polynomials.  Therefore, in the lowest order where $l=0$ and in the  non-relativistic limits  $q^0=m_{_{V}}$ and $(m^2_{_{V}}-m^2_{_{V'}})^2\gg4(k.q)^2$ one has
\begin{eqnarray} \label{I1200}
\dot{I}({\bf{k}})&=& 0\\
\label{Q1200}
\dot{Q}({\bf{k}})&=&\frac{\pi}{2} \frac{ k^{0}m_{_{V}}\sin^2\alpha}{m_{_{V'}}^2(m^2_{_{V'}}-m^2_{_{V}})}\,(\delta_{1}-\delta_{2})n_{DM}({\bf{x}})\left( CV({\bf{k}}\right)), 
\\ \label{U1200}
\dot{U}({\bf{k}})&=&\frac{\pi}{2}\frac{ k^{0}m_{_{V}}\sin^2\alpha}{m_{_{V'}}^2(m^2_{_{V'}}-m^2_{_{V}})}(\delta_{1}-\delta_{2})n_{DM}({\bf{x}})\left(DV({\bf{k}})\right), 
\\ \label{V1200}
\dot{V}({\bf{k}})&=&-\frac{\pi}{2}\frac{ k^{0}m_{_{V}}\sin^2\alpha}{m_{_{V'}}^2(m^2_{_{V'}}-m^2_{_{V}})}(\delta_{1}-\delta_{2})n_{DM}({\bf{x}})\left( (CQ({\bf{k}})+DU({\bf{k}})\right),\nonumber\\
\end{eqnarray}
where $C=4g'_{_{V}}g_{_{V}}\cos^2\theta_{_{W}}$, $D=(4g'^2_{_{V}} -g_{_{V}}^2)\cos^2\theta_{_{W}}$, the angle $\alpha$ is related to the incoming or outgoing photon and $n_{DM}({\bf{x}})$ is the DM density which is defined as [see appendix A for more detail]
\begin{eqnarray}\label{densityint}
n_{ DM}({\bf{x}})\equiv\int \frac{d^3{\bf{q}}}{(2\pi)^3}\rho^{0}({\bf{x}},{\bf{q}}).
\end{eqnarray}
It should be noted that in (\ref{Q1200}) to (\ref{V1200}) for  $\rho_{11}=\rho_{22}\neq\rho_{33}$  there is not any circular polarization at $l=0$ even though the densities of transverse and  longitudinal modes are different.  Nevertheless, for  the next non-zero order which is  $l=2$ the situation is different.  In this case the polarized VDM-photon forward scattering affect  the time evolution of the Stocks parameters as follows
\begin{eqnarray} \label{I1210}
\dot{I}({\bf{k}})&=& 0\\
\label{Q1210}
\dot{Q}({\bf{k}})&=&-\frac{\pi}{5}\frac{ k^{0}m_{_{V}}\sin^2\alpha}{m_{_{V'}}^2(m^2_{_{V'}}-m^2_{_{V}})}\,(\delta_{1}-\delta_{3})n_{2DM}({\bf{x}})\left( CV({\bf{k}}\right)),
\\ \label{U1210}
\dot{U}({\bf{k}})&=&\frac{\pi}{5}\frac{ k^{0}m_{_{V}}\sin^2\alpha}{m_{_{V'}}^2(m^2_{_{V'}}-m^2_{_{V}})}\,(\delta_{1}-\delta_{3})n_{2DM}({\bf{x}})\left( DV({\bf{k}}\right)),
\\ \label{V1210}
\dot{V}({\bf{k}})&=&\frac{\pi}{5}\frac{ k^{0}m_{_{V}}\sin^2\alpha}{m_{_{V'}}^2(m^2_{_{V'}}-m^2_{_{V}})}(\delta_{1}-\delta_{3}))n_{2DM}({\bf{x}})\left( (CQ({\bf{k}})+DU({\bf{k}})\right),
\end{eqnarray}
where $n_{2 DM}({\bf{x}})$ is the quadrupole  contribution of DM density which is defined as [see appendix A for more detail]
\begin{eqnarray}\label{densityint}
n_{2 DM}({\bf{x}})\equiv\int \frac{d^3{\bf{q}}}{(2\pi)^3}\rho^{0}({\bf{x}},{\bf{q}})P_{2}(\cos\theta),
\end{eqnarray}
in which $\theta$ is measured with respect to the DM direction and $P_{2}(\cos\theta)$ is the Legendre polynomial of rank 2. This means that to generate the CP via VDM-photon forward scattering in $l=2$,  the distribution of VDM must have a quadrupole component in addition to the polarized distribution.  However, as (\ref{I1210})-(\ref{V1210}) show, a small difference in the densities of the  transverse and  longitudinal modes for $l=2$ in contrast with $l=0$ is enough to producing a circular polarization.
Meanwhile, (\ref{I1200})-(\ref{V1200}) and (\ref{I1210})-(\ref{V1210}) can be simplified as
\begin{eqnarray} \label{IQUV1}
\dot{I}({\bf{k}})=0,\,\,\,\,\,\,\,\,\,\,\dot{V}({\bf{k}})\approx -\dot{\kappa}\sin^2\alpha\,\,(Q({\bf{k}})+\eta U({\bf{k}})),\nonumber\\
\dot{Q}({\bf{k}})\approx \dot{\kappa}\sin^2\alpha V({\bf{k}}),\,\,\,\,\,\,\,\,\,\dot{U}({\bf{k}})\approx \dot{\kappa}\sin^2\alpha\,\,\eta V({\bf{k}}), \label{stoke012}
\end{eqnarray}
where  $\eta$ is a number and for $l=0$ one has
\begin{eqnarray}\label{SIGTOT1}
\dot{\kappa}&\sim&\sqrt{\langle\sigma v_{rel}\rangle_{ann}}\frac{ k^{0}}{(m^2_{_{V'}}-m^2_{_{V}})}\,(\delta_{1}-\delta_{2})n_{DM}({\bf{x}}),\nonumber\\&=&\sqrt{\langle\sigma v_{rel}\rangle_{ann}}\frac{ k^{0}}{m_{_{V}}^2(X^2-1)}\,(\delta_{1}-\delta_{2})n_{DM}({\bf{x}}),
\end{eqnarray} 
while for $l=2$
\begin{eqnarray}\label{SIGTOT2}
\dot{\kappa}&\sim&\sqrt{\langle\sigma v_{rel}\rangle_{ann}}\,\,\frac{ k^{0}}{(m^2_{_{V'}}-m^2_{_{V}})}(\delta_{1}-\delta_{3})n_{2DM}({\bf{x}}),\nonumber\\&=&\sqrt{\langle\sigma v_{rel}\rangle_{ann}}\,\,\frac{ k^{0}}{m_{_{V}}^2(X^2-1)}\,(\delta_{1}-\delta_{3})n_{2 DM}({\bf{x}}).
\end{eqnarray}
In the above relations  we have defined

\begin{equation}
X=\frac{m_{_{V'}}}{m_{_{V}}},
\end{equation}
and
\begin{eqnarray}\label{SIGTOT02}
g_{_{V}}^4\cos^4\theta_{_{W}}\sim g_{_{V}}^{'4}\cos^4\theta_{_{W}}\sim g_{_{V}}^{2}g_{_{V}}^{'2}\cos^4\theta_{_{W}}\sim \left\langle \sigma v_{rel} \right\rangle_{ann}\frac{m_{_{V'}}^4}{m_{_{V}}^2}.
\end{eqnarray} 
It should be noted that as the relations in (\ref{IQUV1}) show the B-mode also can be generated through the interaction with the VDM.  Unfortunately, in this case the B-mode power spectrum depends on the circular power spectrum which can be estimated as \cite{B-estimate}
\begin{equation}\label{Bmode}
C^{B(S)}_{l}\propto\overline{ \bigg(\frac{\dot{\kappa}}{\dot{\tau}_{e\gamma}}\bigg)^2} C^{V(S)}_{l},
\end{equation}
that is too small for $\overline{ \bigg(\frac{\dot{\kappa}}{\dot{\tau}_{e\gamma}}\bigg)^2}\ll 1$.  Therefore, at the forward scattering level we only examine the circular polarization of the CMB-photons.
\subsection{DAMPING TERM OF DM-PHOTON INTERACTION}
In the previous subsection the forward scattering term of (\ref{roij}) has been considered.  In this subsection we explore the time evolution of  the Stockes parameters from the smaller damping term.  To evaluate the damping term for DM-photon interaction in the case of a completely unpolarized VDM with the distribution function $\rho^{0}(\bf{x},\bf{q})$, we only consider the largest contribution in the time evolution of Stocks parameters which can be obtained as follows (see appendix B)
\begin{eqnarray} \label{stokeDM1}
&&\dot{Q}({\bf{k}})=-\frac{4}{15}\frac{n_{DM}({\bf{x}})}{m_{_{V}}^{2}}\frac{k^{0^{4}}\sin^2\alpha\cos^4\theta_{_{W}}}{(m_{_{V'}}^{2}-m_{_{V}}^{2})^2}(g_{_{V}}^{4}-16g_{_{V}}^{'4})\,I_{2}({\bf{p}}),
\\ \label{stokeDM2}
&&\dot{U}({\bf{k}})=-\frac{16}{15}\frac{n_{DM}({\bf{x}})}{m_{_{V}}^{2}}\frac{k^{0^{4}}\sin^2\alpha\cos^4\theta_{_{W}}}{(m_{_{V'}}^{2}-m_{_{V}}^{2})^2}g_{_{V}}g'_{_{V}}(g_{_{V}}^{2}+4g_{_{V}}^{'2})\,I_{2}({\bf{p}}),
\\ \label{stoke1DM3}
&&\dot{V}({\bf{k}})=0,
\end{eqnarray}
where
\begin{eqnarray} \label{I2}
I_{2}({\bf{p}})=\int \frac{d\Omega'}{4\pi}P_{2}(\cos\alpha')I({\bf{p}}),
\end{eqnarray}
and by using  (\ref{SIGTOT02}), one has
\begin{eqnarray}
&&\dot{V}({\bf{k}})= 0,\nonumber\\
&&\dot{Q}({\bf{k}})\approx -\dot{\tau}_{DM} I_2({\bf{{\bf{p}}}}),\,\,\,\,\,\,\,\,\,\dot{U}({\bf{k}})\approx -\dot{\tau}_{DM} I_2({\bf{p}}), \label{stokeDM12}
\end{eqnarray}
with
\begin{eqnarray} \label{tauD}
\dot{\tau}_{DM}\approx\langle \sigma v_{rel}\rangle_{ann}\,n_{DM}({\bf{x}}) \left( \frac{k^0}{m_{_{V}}}\right) ^4.
\end{eqnarray}
 As (\ref{stokeDM12}) shows, there is no  chance to have CP via damping term for the DM-photon interaction with a completely  unpolarized VDM,  whereas the Faraday rotation can be happened if one has a quadruple degree of freedom $I_2({\bf{p}})$  in the photon intensity. However, the factor $ (\frac{k^0}{m_V})^4$ is very small in the case of CMB photon $k^0<1$eV and massive VDM $m_{_{V}}<1$ MeV. In fact, in comparison with the Thomson scattering,  the main interaction of standard scenario of cosmology, one has
\begin{equation}\label{Op0}
\frac{\dot{\tau}_{DM}}{\dot{\tau}_{e\gamma}}\approx\frac{n_{DM}}{n_e}\frac{\langle\sigma v_{rel}\rangle_{ann}}{\sigma_T} \left( \frac{k^0}{m_{_{V}}}\right) ^4,
\end{equation}
where $\dot{\tau}_{e\gamma}=n_e \sigma_T$ is the Thomson scattering optical depth, $n_e$ is the electron number density and $\sigma_T$  is the Thomson cross section. If we use  the approximation $n_{e}=n_{p}\approx n_{B.M}$ that $n_{p}$ and $ n_{B.M}$ are the number density of proton and baryonic matter, respectively, we can write the ratio (\ref{Op0}) as follows
\begin{eqnarray} \label{Op2}
\frac{\dot{\tau}_{DM}}{\dot{\tau}_{e\gamma}}&\approx&10^{-18}\left( \frac{k^0}{m_{_{V}}}\right) ^4  \left(\frac{n_{DM}}{1.4\times10^{-6}1/cm^3}\right)\left(\frac{n_{e}}{2\times10^{-7}1/cm^3}\right)\nonumber\\&\times& \left( \frac{\langle\sigma v_{rel}\rangle_{ann}}{10^{-32}cm^3/s}\right)\left( \frac{\sigma_{T}}{6.6\times10^{-25}cm^2}\right),
\end{eqnarray}  
where for $\frac{k^0}{m_{_{V}}}<<1$ leads to  $\frac{\dot{\tau}_{DM}}{\dot{\tau}_{e\gamma}}<<1$.  Therefore, the damping term can have a considerable contribution for CMB polarization when $\frac{k^0}{m_{_{V}}}\sim 1$ which is not in the context of this paper. 

\section{CMB CIRCULAR POLARIZATION}
In this section, we are going to examine the CMB CP which can be produced by the VDM-photon forward scattering to study the nature of this type of dark matter.
\subsection{ BOLTZMANN EQUATION FOR CMB PHOTONS WITH THE SCALAR PERTURBATION}
In the presence of primordial scalar perturbations (indicated with a subscript $S$) the temperature anisotropy ($\Delta^{(S)}_{I}$) and polarization ($\Delta^{(S)}_{P}$) of CMB radiation can be described in terms of multipole moments in the conformal time $\eta$  as follows \cite{Seljak,Zaldarriaga}
\begin{equation}\label{BE1}
\Delta^{(S)}_{I,P}(K,\mu,\eta)=\sum_{l=0}^{\infty}(2l+1)(-i)^l\Delta^{(S)}_{I,P_{l}}(K,\eta)P_{l} (\mu),
\end{equation}
in which $\mu=\hat{\bf{n}}.\hat{\bf{K}}=\cos\alpha$ where $\alpha$ is the angle between the CMB photon direction $\hat{\bf{n}}=\frac{{\bf{k}}}{|\bf{k}|}$
and the wave vector $\hat{\bf{K}}$. $P_{l}(\mu)$ is the Legendre polynomial of rank $l$.
Meanwhile, the Boltzmann equations for CMB photons due to the Compton and VDM-photon forward scattering in presence of the scalar perturbations are given by
\begin{eqnarray}\label{BE2}
\dfrac{d}{d\eta}\Delta^{(S)}_{I}+ik\mu\Delta^{(S)}_{I} +4[\dot{\varPsi}-ik\mu\Phi]=\dot{\tau}_{e\gamma}\left[ -\Delta^{(S)}_{I}+\Delta^{(S_{0})}_{I}+4\mu v_{b}+\dfrac{1}{2}P_{2}(\mu)\Pi\right],
\end{eqnarray}
\begin{eqnarray}\label{BE3}
\dfrac{d}{d\eta}\Delta^{\pm (S)}_{P}+ik\mu\Delta^{\pm (S)}_{P} &=&\dot{\tau}_{e\gamma}\left[-\Delta^{\pm (S)}_{P}-\dfrac{1}{2}[1-P_{2}(\mu)]\Pi \right]\nonumber\\
&+& ia(\eta) \dot{\kappa}_{DM}(D\mp iC)\Delta^{(S)}_{V},
\end{eqnarray}
\begin{eqnarray}\label{BE3}
\dfrac{d}{d\eta}\Delta^{(S)}_{V}+ik\mu\Delta^{(S)}_{V}& =&\dot{\tau}_{e\gamma}\left[ -\Delta^{(S)}_{V}+\dfrac{3}{2}\mu\Delta^{(S)}_{V1} \right]\nonumber\\
&+&\frac{i}{2}\dot{\kappa}_{DM}\bigg((D+ iC)\Delta^{-(S)}_{P}-(D- iC)\Delta^{+ (S)}_{P}\bigg),
\end{eqnarray}
 where
\begin{eqnarray} \label{U12}
\dot{\kappa}_{DM}&=&\dot{\kappa}\sin^2\alpha,
\end{eqnarray}
$\dot{\tau}_{e\gamma}\equiv\frac{d\tau_{e\gamma}}{d\eta}$, $a(\eta)$ is normalized scale factor and $\Pi\equiv\Delta_{I_{2}}^{(S)}+\Delta_{P_{2}}^{(S)}-\Delta_{P_{0}}^{(S)}$. Also, $\Psi$ and $\Phi$ are metric  perturbations which   correspond to the Newtonian potential and the perturbation to the spatial curvature, respectively.
The values of $\Delta^{\pm(S)}_{P}(\hat{\bf{n}})=Q^S(\hat{\bf{n}})\pm U^S(\hat{\bf{n}})$ and $\Delta^{(S)}_{V}(\hat{\bf{n}})=V^S(\hat{\bf{n}})$ at the present time $\eta_{0}$ and the direction $\hat{\bf{n}}$ are given in the following general
form by integrating of the Boltzmann equations (\ref{BE1})-(\ref{BE3}) along the line of sight \cite{Seljak} and summing over all the Fourier modes $\bf{K}$ as follows
\begin{eqnarray}
\Delta^{\pm(S)}_{P}(\hat{\bf{n}})=\int d^3{\bf{k}}\,\xi({\bf{K}})e^{\pm 2i\psi_{K,n}}\Delta^{\pm(S)}_{P}({\bf{K}},\mu,\eta_{0}),
\end{eqnarray}
\begin{eqnarray}
\Delta^{(S)}_{V}(\hat{\bf{n}})=\int d^3{\bf{k}}\,\xi({\bf{K}})\Delta^{(S)}_{V}({\bf{K}},\mu,\eta_{0}),
\end{eqnarray}
where $\psi_{K,n}$ is the angle needed to rotate the $\bf{K}$ and $\hat{\bf{n}}$ dependent basis to a fixed frame in the sky, $\xi(\bf{k})$ is a random
variable using to characterize the initial amplitude of each primordial scalar perturbations mode, and also the values
of  $\Delta^{\pm(S)}_{P}({\bf{K}},\mu,\eta_{0})$ and $\Delta^{(S)}_{V}({\bf{K}},\mu,\eta_{0})$ are given as
\begin{equation}\label{delp}
\Delta^{\pm (S)}_{P}({\bf{K}},\mu,\eta_{0})=\int_{0}^{\eta_{0}} d\eta\,\dot{\tau}_{e\gamma}\, e^{ix\mu-\tau_{e\gamma}}\,\bigg[\dfrac{3}{4}(1-\mu^{2})\,\Pi(K,\eta)+ i\frac{\dot{\kappa}_{DM}}{\dot{\tau}_{e\gamma}}(D\mp iC)\Delta^{(S)}_{V}\bigg],
\end{equation}

\begin{eqnarray}\label{delv}
\Delta^{ (S)}_{V}({\bf{K}},\mu,\eta_{0})&=&\frac{1}{2}\int_{0}^{\eta_{0}} d\eta\,\dot{\tau}_{e\gamma}\, e^{ix\mu-\tau_{e\gamma}}\,\bigg[3\mu\Delta^{(S)}_{V1} \nonumber\\
&+&i\frac{\dot{\kappa}_{DM}}{\dot{\tau}_{e\gamma}}\bigg((D+ iC)\Delta^{- (S)}_{P}-(D- iC)\Delta^{+ (S)}_{P}\bigg)\bigg],\\
&\approx&\frac{1}{2}\int_{0}^{\eta_{0}} d\eta\,\dot{\tau}_{e\gamma}\, e^{ix\mu-\tau_{e\gamma}}\,\bigg[3\mu\Delta^{(S)}_{V1} +2iC\frac{\dot{\kappa}_{DM}}{\dot{\tau}_{e\gamma}}\Delta^{(S)}_{P}\bigg],
\end{eqnarray}
where
\begin{equation}\label{delp2}
\Delta^{(S)}_{P}({\bf{K}},\mu,\eta_{0})=\int_{0}^{\eta_{0}} d\eta\,\dot{\tau}_{e\gamma}\, e^{ix\mu-\tau_{e\gamma}}\,\bigg[\dfrac{3}{4}(1-\mu^{2})\,\Pi(K,\eta)\bigg],
\end{equation}
and $x=K(\eta_{0}-\eta)$. In the above equations the differential optical depth $\dot{\tau}_{e\gamma}(\eta)$ and the total optical depth $\tau_{e\gamma}(\eta)$ due to the Compton scattering at time $\eta$ are given as
\begin{equation}\label{BE7}
\dot{\tau}_{e\gamma}(\eta)=a(\eta)n_{e}\sigma_{T},\,\,\,\,\,\tau_{e\gamma}(\eta)=\int_{\eta_{0}}^{\eta}\dot{\tau}_{e\gamma}(\eta)\,d\eta.
\end{equation}
\subsection{THE  POWER	SPECTRUM OF CMB CIRCULAR POLARIZATION}
\begin{figure}
\centering
\includegraphics[width=4in]{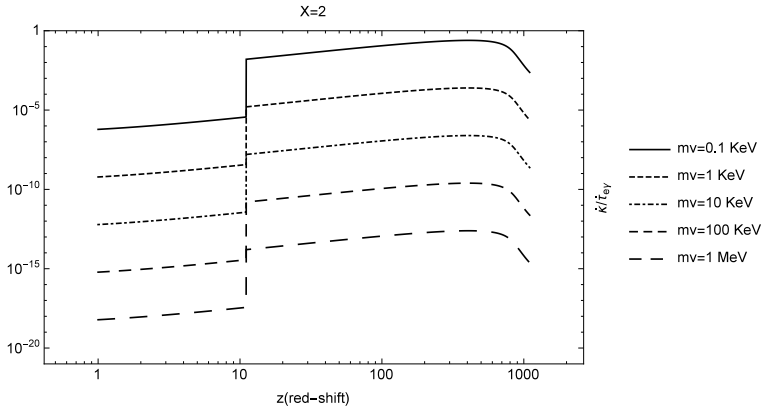}
\caption{ Optical depths ratio $\frac{\dot{\kappa}}{\dot{\tau}_{e\gamma}}$ for $l=0$ in terms of red-shift for different  masses $m_V\sim 0.1keV,\,1keV,\,10keV,\,100keV,\,1MeV$ with $X=m_{V'}/m_V=2$ and $\langle\sigma v_{rel}\rangle=10^{-32}{\rm cm^3 s^{-1}}$.}\label{Fig2}
\end{figure}

\begin{figure}
\centering
\includegraphics[width=4in]{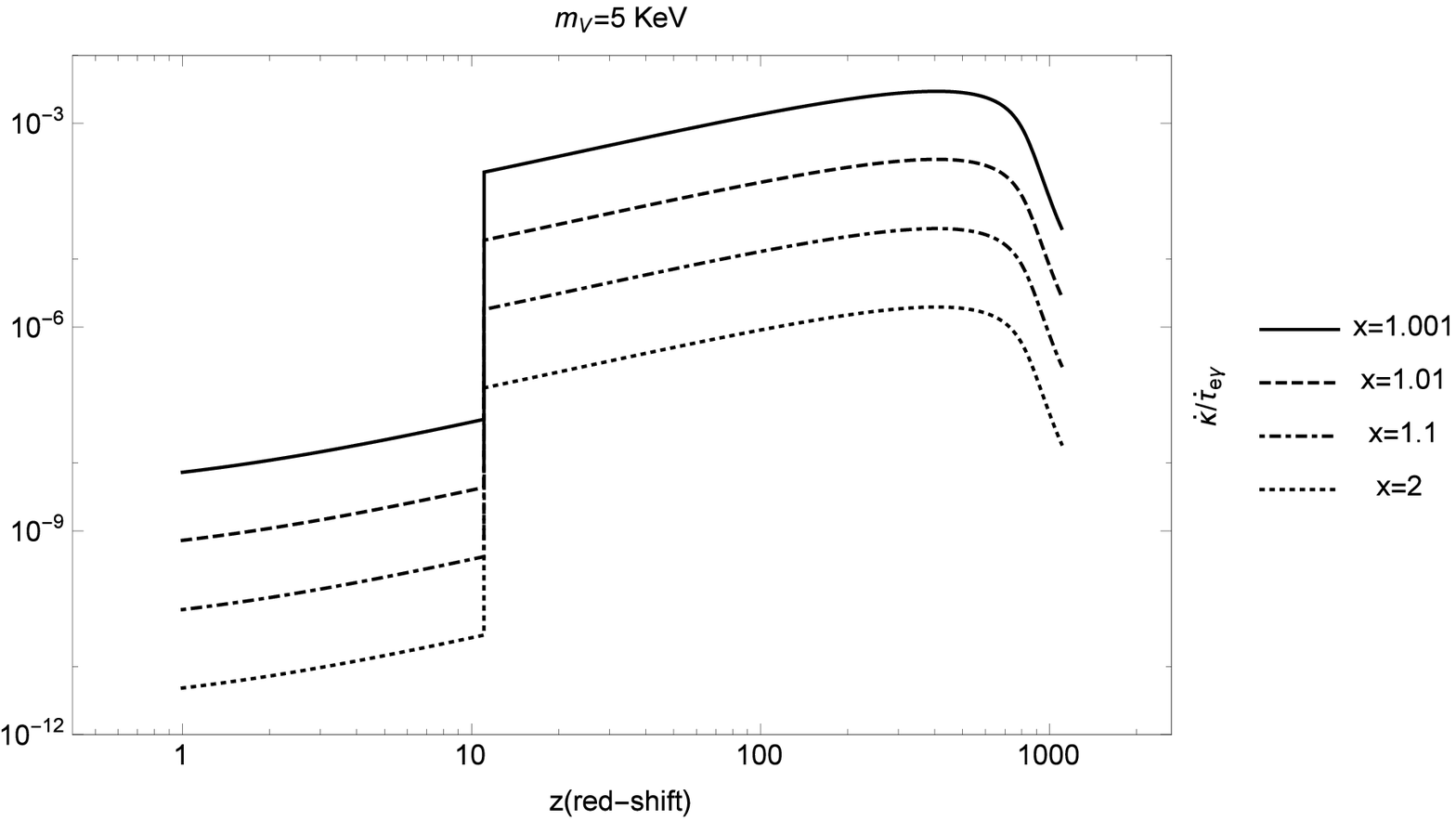}
\caption{Optical depths ratio $\frac{\dot{\kappa}}{\dot{\tau}_{e\gamma}}$ for $l=0$  in terms of red-shift for mass $m_V\sim 5keV$ with different mass ratios $X=m_{_{V'}}/m_{_{V}}\sim 2, 1.1, 1.01, 1.001$ and $\langle\sigma v_{rel}\rangle=10^{-32}{\rm cm^3 s^{-1}}$.}\label{Fig3}
\end{figure}

\begin{figure}[htb]
\centering
\begin{tabular}{@{}cccc@{}}
\includegraphics[width=.5\textwidth]{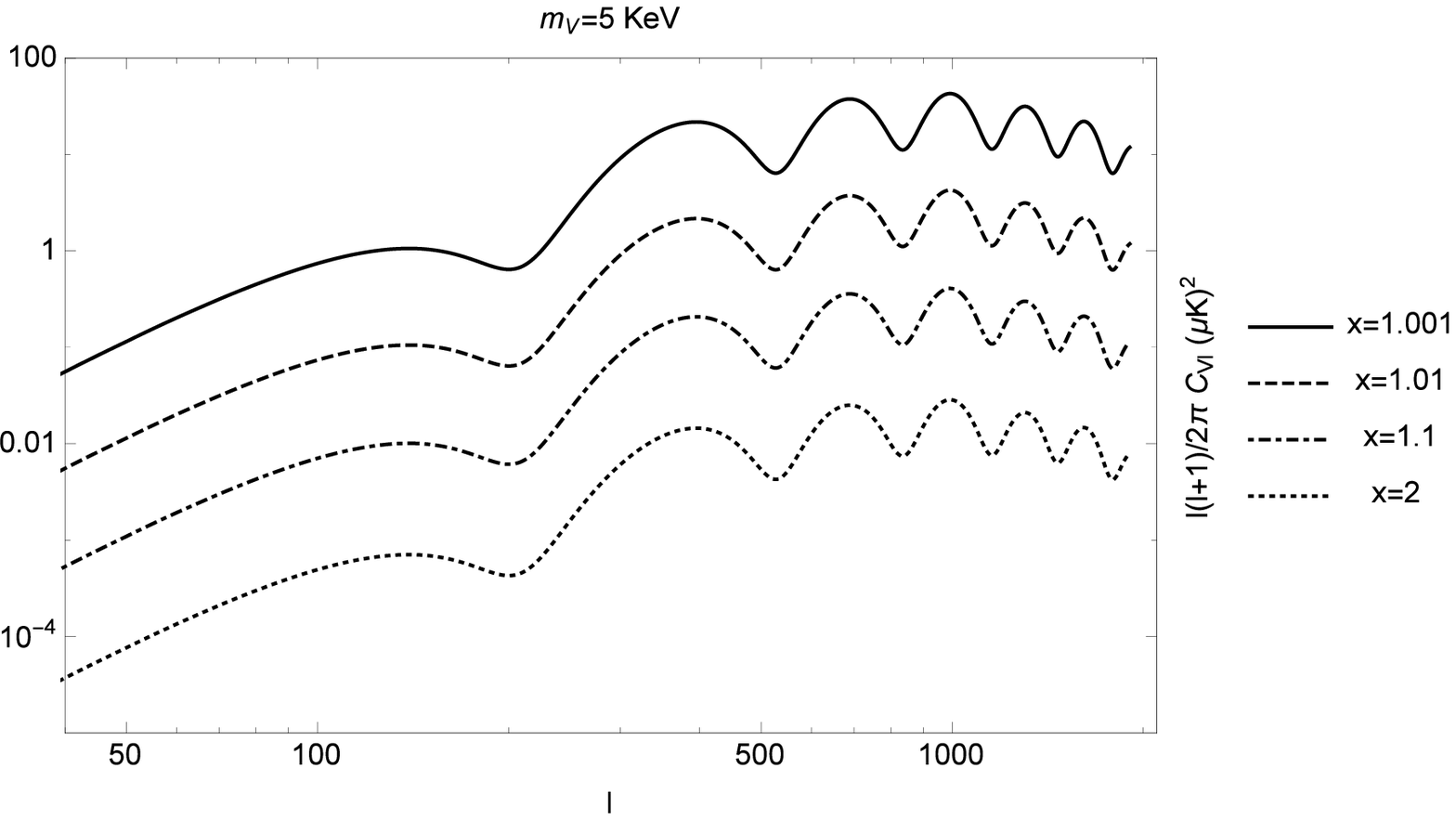} &	\includegraphics[width=.5\textwidth]{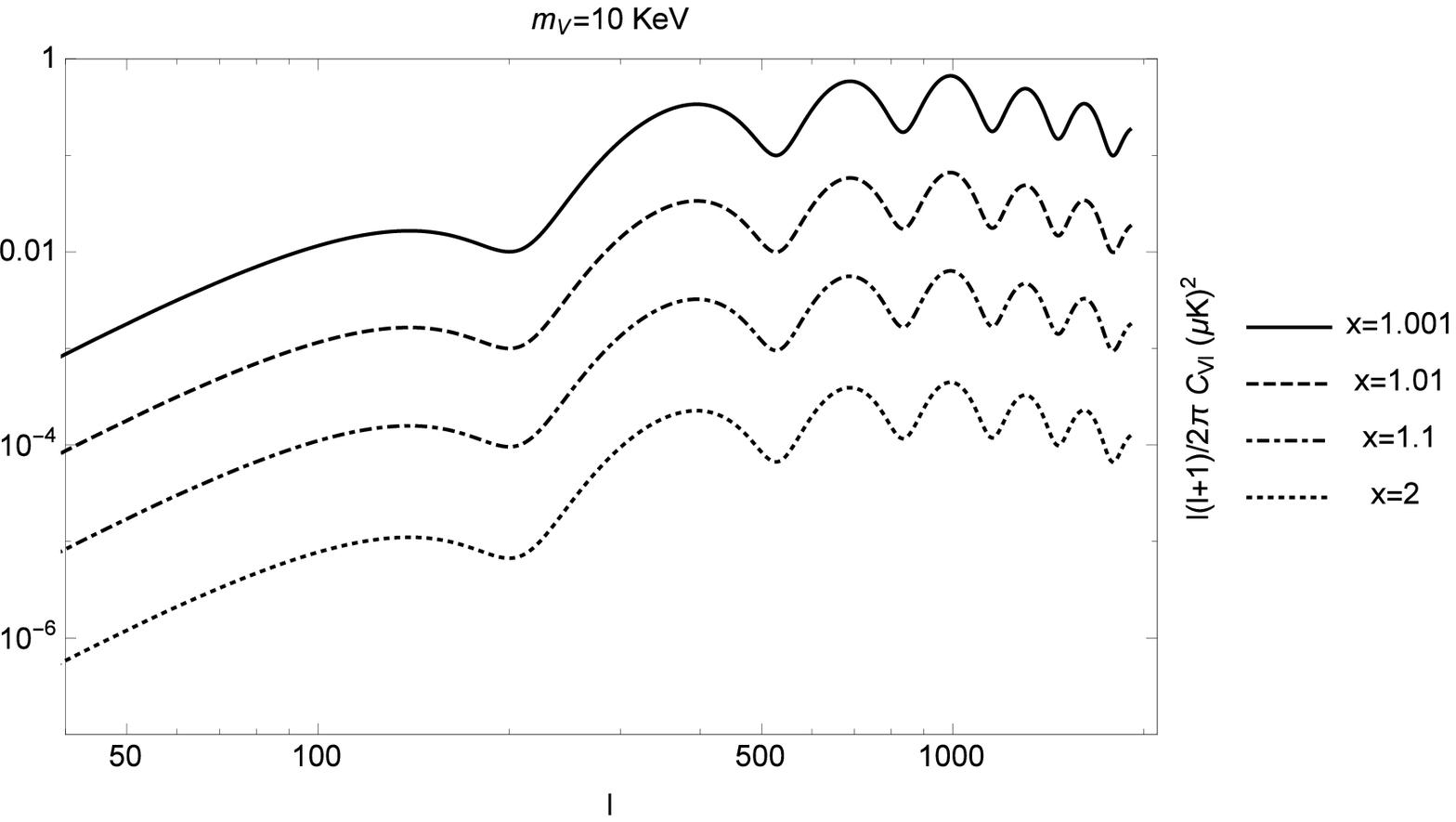}    \\
\multicolumn{2}{c}{\includegraphics[width=.5\textwidth]{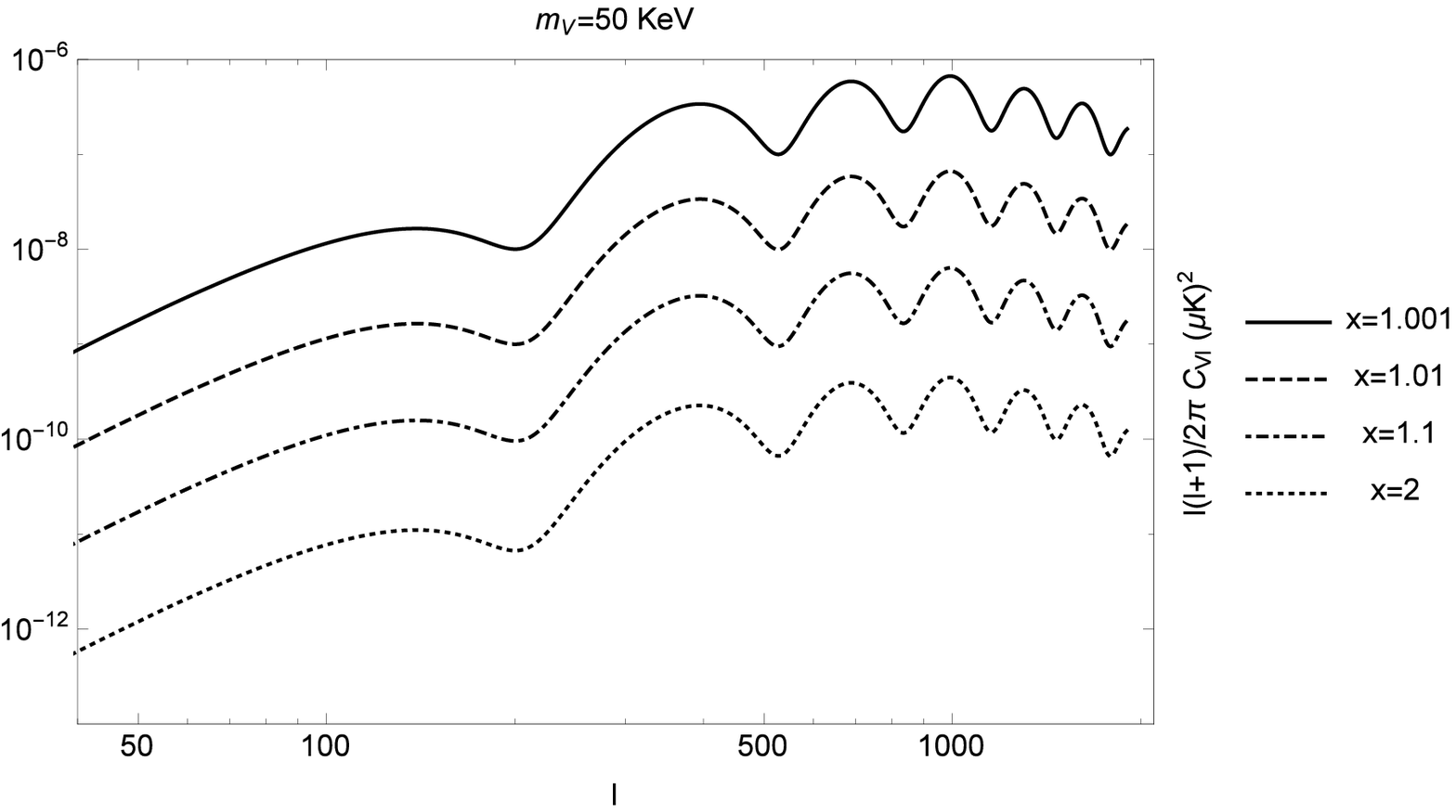}}
\end{tabular}
\caption{CP angular power spectrum $l (l+1)/(2\pi)\,C^{(S)}_{Vl}$ for $l=0$  in terms of the angular momentum $l$ for $m_{_{V}}\sim 5keV, 10keV, 50keV$ with different mass ratios $X=m_{_{V'}}/m_{_{V}}\sim 2, 1.1, 1.01, 1.001$ and $\langle\sigma v_{rel}\rangle=10^{-32}{\rm cm^3 s^{-1}}$.}\label{Fig4}
\end{figure}

\begin{figure}
\centering
\includegraphics[width=4in]{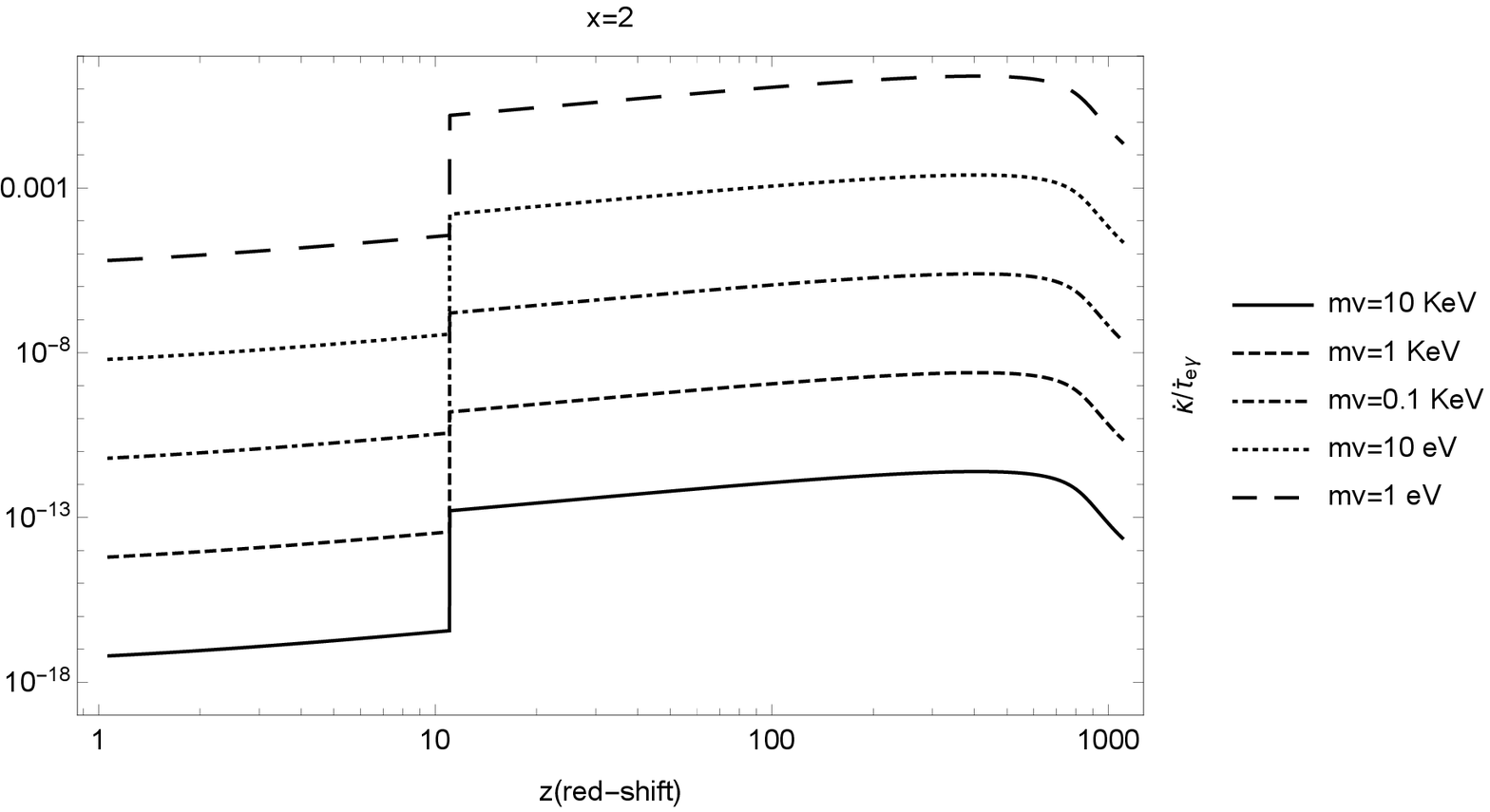}
\caption{ Optical depths ratio $\frac{\dot{\kappa}}{\dot{\tau}_{e\gamma}}$ for $l=2$  in terms of red-shift for different  masses $m_V\sim 1eV,\,10eV,\,0.1keV,\,1keV,\,10keV$ with $X=m_{V'}/m_V=2$ and $\langle\sigma v_{rel}\rangle=10^{-32}{\rm cm^3 s^{-1}}$.}\label{Fig5}
\end{figure}

\begin{figure}[htb]
\centering
\begin{tabular}{@{}cccc@{}}
\includegraphics[width=.45\textwidth]{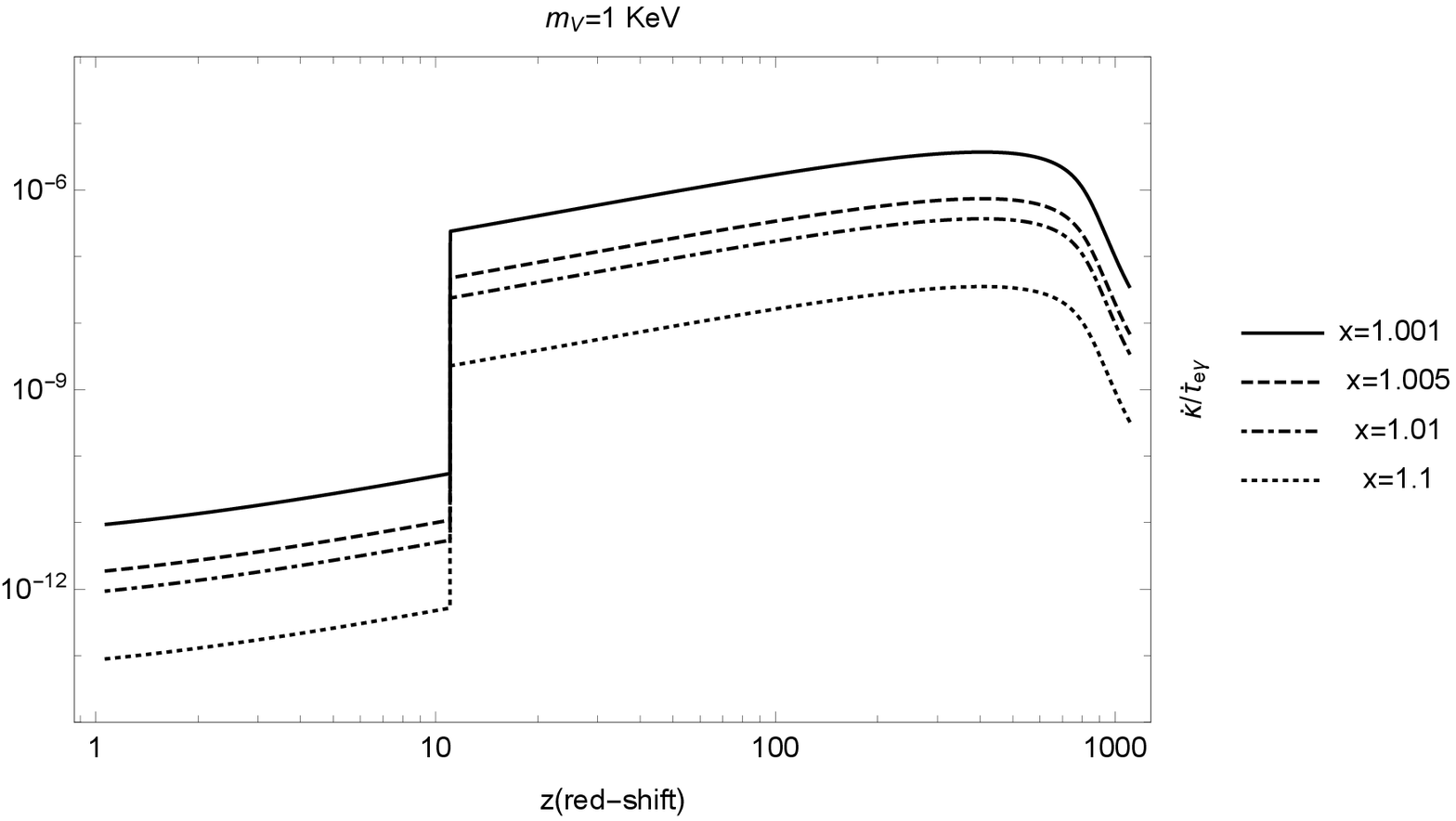} &
\includegraphics[width=.45\textwidth]{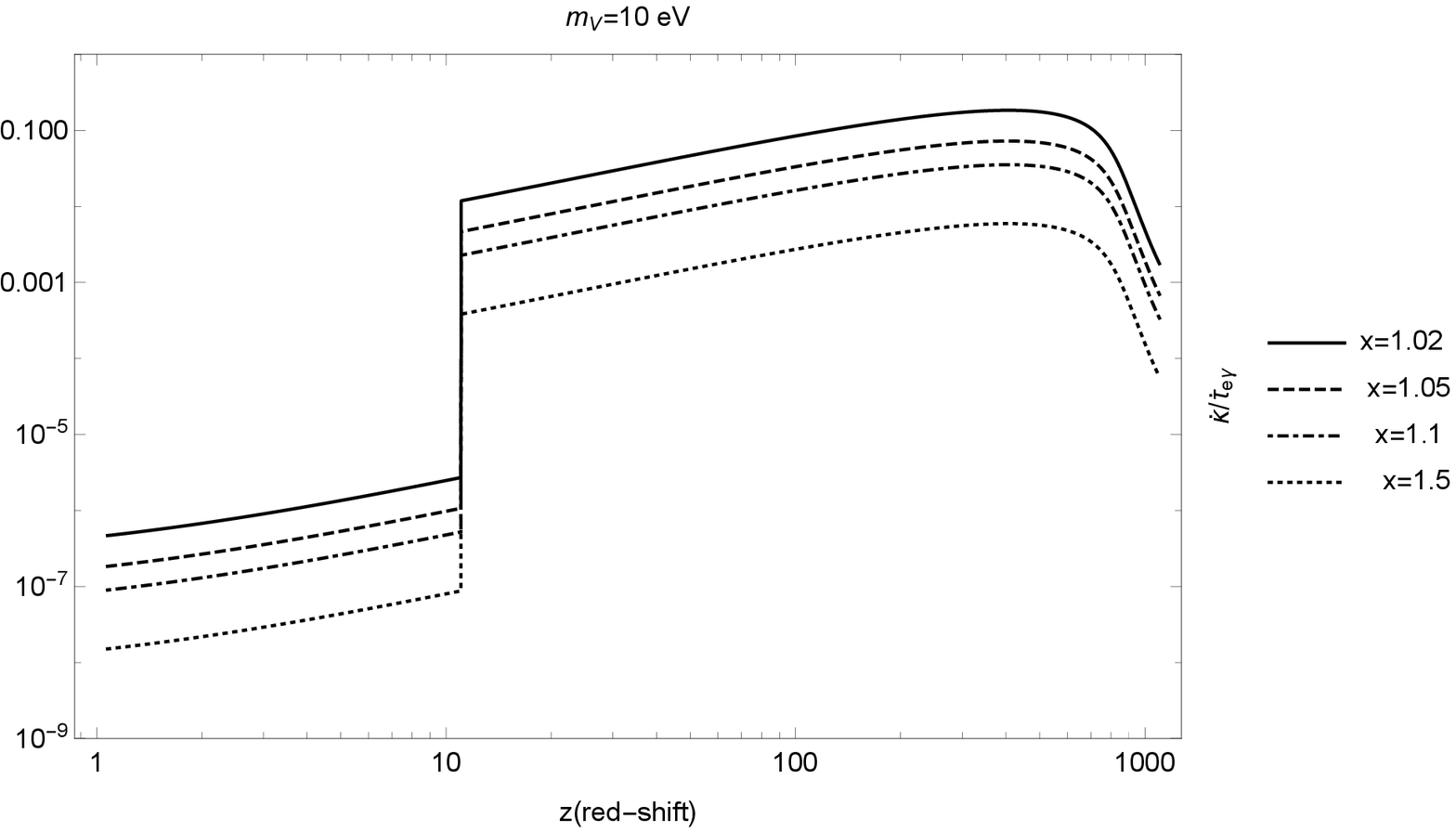}    
\end{tabular}
\caption{Optical depths ratio $\frac{\dot{\kappa}}{\dot{\tau}_{e\gamma}}$ for $l=2$  in terms of red-shift for masses $m_V\sim 1keV,10eV$ with different mass ratios $X=m_{_{V'}}/m_{_{V}}$ and $\langle\sigma v_{rel}\rangle=10^{-32}{\rm cm^3 s^{-1}}$.}\label{Fig6}
\end{figure}

\begin{figure}[htb]
\centering
\begin{tabular}{@{}cccc@{}}
\includegraphics[width=.45\textwidth]{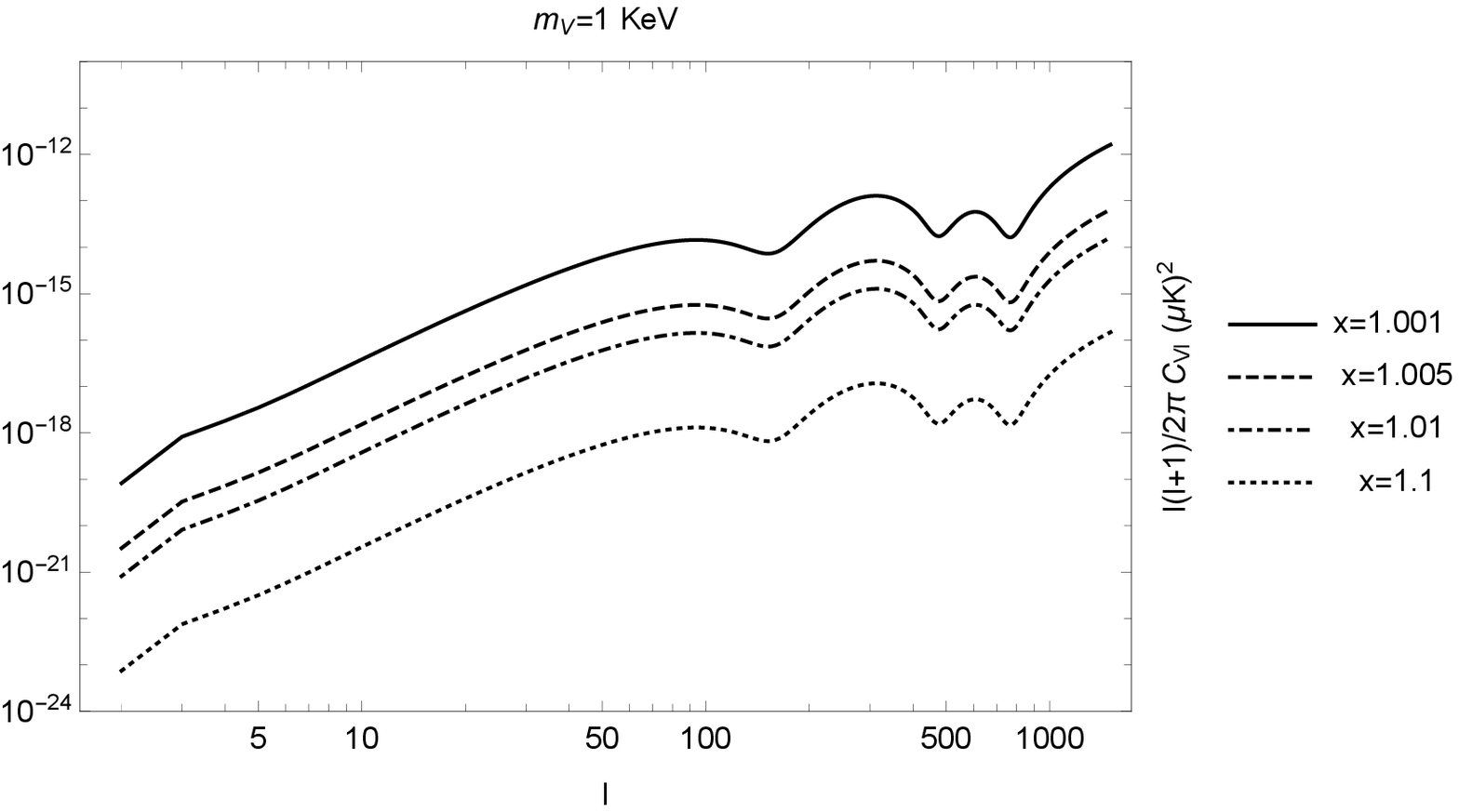} &
\includegraphics[width=.45\textwidth]{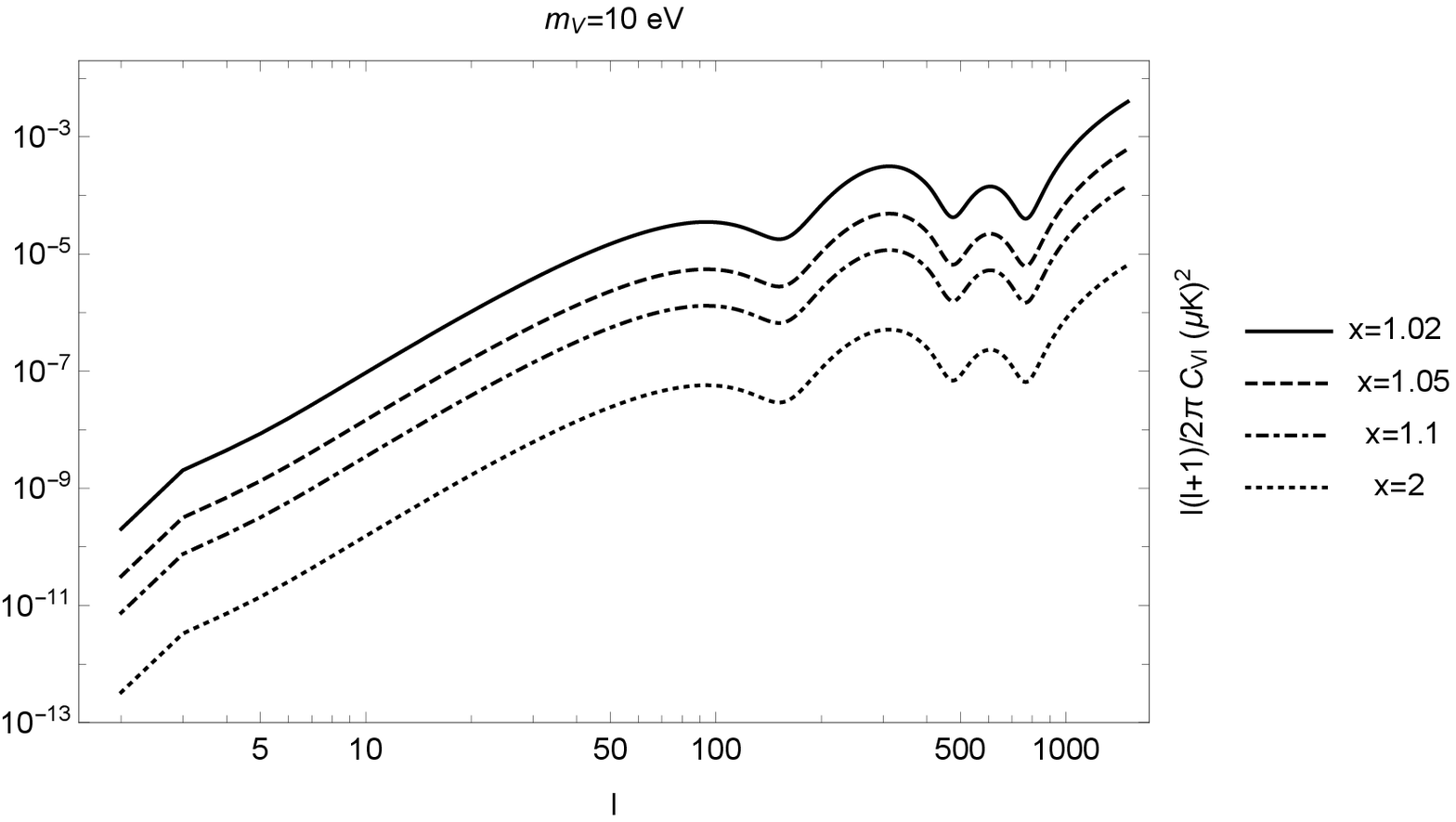}    
\end{tabular}
\caption{CP angular power spectrum $l (l+1)/(2\pi)\,C^{(S)}_{Vl}$ for $l=2$   in terms of the angular momentum $l$ for $m_{_{V}}\sim 1keV$ with different mass ratios $X=m_{_{V'}}/m_{_{V}}$ and $\langle\sigma v_{rel}\rangle=10^{-32}{\rm cm^3 s^{-1}}$.}\label{Fig7}
\end{figure}

Now we are ready to calculate the power spectra $C_{Vl}^{(S)}$ of CP of CMB
radiation due to the Compton scattering and VDM-photon forward scattering in the presence of scalar perturbations.
In the presence of primordial scalar perturbation, the power spectrum $C_{Vl}^{(S)}$ is given by
\begin{equation}\label{ClV}
C_{Vl}^{(S)}=\frac{1}{2l+1}\sum_{m}\left\langle a^{\ast}_{V,lm} a_{V,lm} \right\rangle,
\end{equation}
where
\begin{eqnarray}\label{ClV1}
a_{V,lm}&=&\int\,d\Omega\,Y^{\ast}_{lm}\Delta_{V}.
\end{eqnarray}
By using (\ref{delp})-(\ref{delp2}), the power spectrum $C^{(S)}_{Vl}$ can be determined as
\begin{eqnarray}\label{ClV2}
C^{(S)}_{Vl}&\approx&\frac{1}{2l+1}\int\,d^3{\bf{K}}\, P_{\phi}({\bf{K}})\nonumber\\
&\times&\sum_{m}
\bigg|\int\,d\Omega\,Y^{\ast}_{lm}\,\int^{\eta_{0}}_{0}\,d\eta\,\dot{\tau}_{e\gamma}\,e^{ix\mu-\tau_{e\gamma}}\,
\frac{\dot\kappa}{\dot{\tau}_{e\gamma}}\,(1-\mu^2)\,\Delta^{(S)}_{P}\bigg|^2,
\end{eqnarray}
with
\begin{eqnarray}\label{ClV3}
P_{\phi}({\bf{K}})\delta({\bf{K'}-\bf{K}})=\bigg\langle\xi(\bf{K})\xi(\bf{K'})\bigg\rangle,
\end{eqnarray}
in which $P_{\phi}({\bf{K}})$ is the primordial scalar spectrum and by using  (\ref{SIGTOT1})and (\ref{SIGTOT2}) , we have $2C\frac{\dot{\kappa}_{DM}}{\dot{\tau}_{e\gamma}}\approx \frac{\dot{\kappa}}{\dot{\tau}_{e\gamma}}(1-\mu^2)$ where for $l=0$

\begin{equation}\label{kappaf1}
\frac{\dot{\kappa}}{\dot{\tau}_{e\gamma}}\sim\frac{(\delta_{1}-\delta{2})}{\alpha}\frac{m_e k^0}{m_{_{V}}^2-m_{_{V'}}^2}\frac{n_{DM}}{n_e}\sqrt{\frac{\left\langle \sigma v_{rel}\right\rangle_{ann}}{\sigma_T}},
\end{equation}
and for $l=2$
\begin{equation}\label{kappaf2}
\frac{\dot{\kappa}}{\dot{\tau}_{e\gamma}}\sim\frac{(\delta_{1}-\delta{3})}{\alpha}\frac{m_e k^0}{m_{_{V}}^2-m_{_{V'}}^2}\frac{n_{2DM}}{n_e}\sqrt{\frac{\left\langle \sigma v_{rel}\right\rangle_{ann}}{\sigma_T}}.
\end{equation}
Meanwhile,  the small deviation $\delta_r$ in the densities from the unpolarized distribution of the VDM  and the ratio $n_{2DM}/n_{DM}$ can be assumed to be about the temperature fluctuation as \cite{tttv,Bennett, pol}
\begin{equation}\label{delta1}
n_{2DM}/n_{DM}\leq\Delta T/T\sim 10^{-5},~~~~~~~~~~~(\delta_{1}-\delta{2}),(\delta_{1}-\delta{3})\leq\Delta T/T\sim 10^{-5}.
\end{equation}
As (\ref{kappaf1}) and  (\ref{kappaf2}) show, the ratio $\frac{\dot{\kappa}}{\dot{\tau}_{e\gamma}}$ is small if the VDM masses are large and far from each other.  Therefore, the cases with the small VDM masses where the masses are near to each other lead to larger values for the CP generation due to the VDM-photon scattering.  Nevertheless, the power spectrum $C_{Vl}^{(S)}$ of the CP depends on the $\frac{\dot{\kappa}}{\dot{\tau}_{e\gamma}}$ through a complicated integral given in (\ref{ClV2}).  However, the effects of different VDM masses on the optical depths ratio $\frac{\dot{\kappa}}{\dot{\tau}_{e\gamma}}$ and the CP angular power spectrum $l (l+1)/(2\pi)\,C^{(S)}_{Vl}$ are shown in  Figs. \ref{Fig2} to \ref{Fig7}.
As the optical depths ratio $\frac{\dot{\kappa}}{\dot{\tau}_{e\gamma}}$ for different masses in Figs. \ref{Fig2}, \ref{Fig3}, \ref{Fig5} and \ref{Fig6} show, impact of the VDM-photon scattering on the CP generation for masses larger than $10keV$ in $l=0$ and for  the masses larger than $1 keV$ in $l=2$ are negligible.  On the other hand,  as the Figs.  \ref{Fig4} and  \ref{Fig7} show the power spectrum increases very fast by reducing the VDM masses as  $\sim  1/m_{_{V}}^6$ which leads to a stringent bound on the low mass VDM. 
For instance, as the Fig. \ref{Fig7}  shows for  $m_{_{V}}\sim10eV$ and  $X=m_{V'}/m_V=1.02$, the CP power spectrum is about $10^{-3}{\rm \mu K^2}=10^3{\rm nK^2}$ which is very larger than $nK^2$ region where one expects for the CP of the CMB.
\newpage


\section{CONCLUSION}
We considered a spin-1 bosonic DM with a direct coupling with photon as is given in (\ref{LI}) .  Consequently, we have explored  the time evolution of the Stocks  parameters to find our results as follows:\\
1-The forward scattering  of an unpolarized VDM with a completely uniform distribution function can not generate any CMB polarization.\\
2-The Damping term for the  unpolarized VDM is not zero but it is negligible in comparison with the  Compton scattering.  In fact, the optical depth of this scattering is much smaller than the optical depth of Compton scattering $\frac{\dot{\tau}_{DM}}{\dot{\tau}_{e\gamma}}\propto(\frac{k^{0}}{m_{_{V}}})^4\ll 1$ where the photon energy $k^0$ is much smaller than the mass of dark matter.\\
3-However, without considering a specific model or mechanism for producing a vector dark mater, we have considered a partially polarized Vector-DM by taking different densities for each components of the vector dark mater as is given in (\ref{matrisro1}).  We found that at the lowest order of the multipole expansion ( ie $l=0$ ) the CP can be generated only if $\rho_{11}\neq\rho_{22}$.  In this case, if exists, the VDM's with masses less than a few $keV$ lead to a large circular polarization for the CMB ( see Fig. \ref{Fig4}) and can be excluded from the acceptable range of the VDM masses.  Nevertheless, in the next leading order for a partially polarized Vector-DM with a quadrupole distribution of the order of temperature fluctuation one has also a valuable CP for the CMB.  Such a new way for producing the CMB polarization has provided a new constrain on the properties of DM particles. \\
4-As one can see from (\ref{kappaf1}) and  (\ref{kappaf2}), to have a valuable impact on the CP of the CMB, one should consider small masses for the VDM's  where the masses are near to each other.  \\
5-We also showed that the CP angular power spectrum depends on the mass of the VDM as $C^{(S)}_{Vl}\propto 1/m_{_{V}}^6$.  Therefore,  the power spectrum decreases very fast by increasing the VDM mass such that for $m_V\gg 1keV$ there is not any measurable CP through the VDM-photon scattering.  \\
6- In different DM-models the DM-mass ranges from a few $meV$ to a few $TeV$.   The $ 1/m_{_{V}}^6$ mass dependence of the power spectrum can put a stringent bound on the low mass VDM.  As the Fig.  \ref{Fig7} for $m_{_{V}}=10eV$  shows for the large angular momentum ($l> 50$) the CP increases to thousands of $nK^2$ by reducing the VDM mass to a few electron volts.  Therefore, the VDMs with  $m_V\lesssim 10 eV$ can be excluded in our model.  For instance, for $m_{_{V}}=10eV-1keV$ the CP angular power spectrum is  $C^{(S)}_{Vl}\simeq 10^3- 10^{-11}{\rm nK^2}$ for $l\simeq 300$.

\section{APPENDIX A: THE TIME EVALUATION OF PHOTON DENSITY MATRIX  DUE TO DM-PHOTON SCATTERING}
In this Appendix we are going to calculate contribution of the VDM-photon forward scattering to the time evolution of the photon density matrix elements in detail. The second-order scattering matrix is given by
\begin{eqnarray}
S^{(2)}=-\frac{1}{2}\int_{-\infty}^{\infty}dt\int_{-\infty}^{\infty}dt'\,T\{H^{(1)}_{int}(t)H^{(1)}_{int}(t')\}\equiv-i\int_{-\infty}^{\infty}dt\, H^{(2)}_{int}(t),
\end{eqnarray}
where $H^{(1)}_{int}(t)$ is the first-order interaction Hamiltonian. From the Lagrangian (\ref{LI}), one can obtain the interaction Hamiltonian for the second-order scattering processes as follows
\begin{eqnarray}
H^{(2)}_{int}(t)&=&-\frac{i}{2}\int\,d^{3}{\bf{x}}\,d^{4}x'\,\nonumber\\
&\times& T\bigg\{\bigg(g_{_{V}}\cos \theta_{_{W}}F^{\mu\upsilon}V_{\mu}V'_{\nu}+ g'_{_{V}}\cos \theta_{_{W}}\varepsilon^{\mu\nu\alpha\beta}F_{\mu\upsilon}V_{\alpha}V'_{\beta}\bigg)_{x}\nonumber\\
&\times&\bigg(g_{_{V}}\cos \theta_{_{W}}F^{\rho\sigma}V_{\rho}V'_{\sigma}+ g'_{_{V}}\cos \theta_{_{W}}\varepsilon^{\rho\sigma\lambda\eta}F_{\rho\sigma}V_{\lambda}V'_{\eta}\bigg)_{x'}\bigg\}.
\end{eqnarray}	
The part of $H^{(2)}_{int}(t)$ which describes the VDM-photon scattering by $H^{0}_{I}(t)$ as
\begin{eqnarray}\label{HI}
H^{0}_{I}(t)\equiv H^{(1)}_{I}(t)+H^{(2)}_{I}(t)+H^{(3)}_{I}(t),
\end{eqnarray}	
where $H^{(1)}_{I}(t)$, $H^{(2)}_{I}(t)$ and $H^{(3)}_{I}(t)$ are
\begin{eqnarray}\label{HI1}
H^{(1)}_{I}(t)&=&-\frac{i}{2}g^2_{_{V}}\cos ^2\theta_{_{W}}\int\,d^{3}{\bf{x}}\,d^{4}x'\, iD_{F}^{\nu\sigma}(x-x')
\nonumber\\
&\times&V^{\mu-}(x)\bigg[\partial_{\mu}A_{\nu}^{-}(x)\partial_{\rho}A_{\sigma}^{+}(x')+\partial_{\rho}A_{\sigma}^{-}(x')\partial_{\mu}A_{\nu}^{+}(x)\nonumber\\
&-&\partial_{\nu}A_{\mu}^{-}(x)\partial_{\rho}A_{\sigma}^{+}(x')-\partial_{\rho}A_{\sigma}^{-}(x')\partial_{\nu}A_{\mu}^{+}(x)\nonumber\\
&-&\partial_{\mu}A_{\nu}^{-}(x)\partial_{\sigma}A_{\rho}^{+}(x')-\partial_{\sigma}A_{\rho}^{-}(x')\partial_{\mu}A_{\nu}^{+}(x)\nonumber\\
&+&\partial_{\nu}A_{\mu}^{-}(x)\partial_{\sigma}A_{\rho}^{+}(x')+\partial_{\sigma}A_{\rho}^{-}(x')\partial_{\nu}A_{\mu}^{+}(x)\bigg]V^{\nu+}(x'),
\end{eqnarray}	
\begin{eqnarray}\label{HI2}
H^{(2)}_{I}(t)&=& -ig_{_{V}}g'_{_{V}}\cos ^2\theta_{_{W}}\varepsilon^{\rho\sigma\lambda\eta}\int\,d^{3}{\bf{x}}\,d^{4}x'\,  iD_{F\nu\eta}(x-x')\nonumber\\
&\times& \bigg(V^{-}_{\mu}(x)
\bigg[\partial^{\mu}A^{\nu-}(x)\partial_{\rho}A_{\sigma}^{+}(x')+\partial_{\rho}A_{\sigma}^{-}(x')\partial^{\mu}A^{\nu+}(x)\nonumber\\
&-&\partial^{\nu}A^{\mu-}(x)\partial_{\rho}A_{\sigma}^{+}(x')-\partial_{\rho}A_{\sigma}^{-}(x')\partial^{\nu}A^{\mu+}(x)\bigg]V^{+}_{\lambda}(x')\nonumber\\
&+&V^{-}_{\lambda}(x)
\bigg[\partial_{\rho}A_{\sigma}^{-}(x)\partial^{\mu}A^{\nu+}(x')+\partial^{\mu}A^{\nu-}(x')\partial_{\rho}A_{\sigma}^{+}(x)\nonumber\\
&-&\partial_{\rho}A_{\sigma}^{-}(x)\partial^{\nu}A^{\mu+}(x')-\partial^{\nu}A^{\mu-}(x')\partial_{\rho}A_{\sigma}^{+}(x)\bigg]V^{+}_{\mu}(x')\bigg),
\end{eqnarray}	
\begin{eqnarray}\label{HI3}
H^{(3)}_{I}(t)&=&-\frac{i}{2}g'^2_{_{V}}\cos ^2\theta_{_{W}}\varepsilon^{\mu\nu\alpha\beta}\varepsilon^{\rho\sigma\lambda\eta}\int\,d^{3}{\bf{x}}\,d^{4}x'\,  iD_{F\beta\eta}(x-x')\nonumber\\
& \times&
V^{-}_{\alpha}(x)\bigg[ \partial_{\mu}A_{\nu}^{-}(x)\partial_{\rho}A_{\sigma}^{+}(x')+\partial_{\rho}A_{\sigma}^{-}(x')\partial_{\mu}A_{\nu}^{+}(x)\nonumber\\
&-&\partial_{\nu}A_{\mu}^{-}(x)\partial_{\rho}A_{\sigma}^{+}(x')
-\partial_{\rho}A_{\sigma}^{-}(x')\partial_{\nu}A_{\mu}^{+}(x)\nonumber\\
&-&\partial_{\mu}A_{\nu}^{-}(x)\partial_{\sigma}A_{\rho}^{+}(x')-\partial_{\sigma}A_{\rho}^{-}(x')\partial_{\mu}A_{\nu}^{+}(x)\nonumber\\
&+&\partial_{\nu}A_{\mu}^{-}(x)\partial_{\sigma}A_{\rho}^{+}(x')+\partial_{\sigma}A_{\rho}^{-}(x')\partial_{\nu}A_{\mu}^{+}(x)\bigg]   V^{+}_{\lambda}(x').
\end{eqnarray}	
In (\ref{HI1})-(\ref{HI3}) $V^{-}(V^{+})$ and $A^{-} (A^{+})$ depend linearly on the creation (absorption) operators of the VDM and photon, respectively. Also, $D^{\alpha\beta}_{F} $ is the Feynman propagator for the VDM.  However, the Fourier transforms of the fields and propagator are given as
\begin{equation}\label{A}
A_{\mu}(x)=\int\frac{d^3 \bf{p}}{(2\pi)^3 2p^0}\sum_{s}\left[ a_{s}(p)\varepsilon_{s\mu}(p)e^{-ip.x} + a_{s}^\dagger (p)\varepsilon_{s\mu}^* (p)e^{ip.x}\right] ,
\end{equation}
\begin{equation}\label{V}
V_{\nu}(x)=\int\frac{d^3\bf{q}}{(2\pi)^3 2q^0}\sum_{r}\left[ b_{r}(q)\varepsilon_{r\nu}(q)e^{-iq.x} + b_{r}^\dagger (q)\varepsilon_{r\nu}^* (q)e^{iq.x}\right] ,
\end{equation}
and
\begin{equation}\label{D}
iD_{F}^{\alpha\beta}(x)=\int \frac{d^4 q}{(2\pi)^4}\frac{i(-g^{\alpha\beta}+q^{\alpha} q^{\beta}/m_{_{V'}}^{2})}{q^2 - m_{_{V'}}^2 + i\theta} e^{-iq.x},
\end{equation}
where $\varepsilon_{s\mu}(k)$ and $ \varepsilon_{r\nu}(q)$, respectively, are the photon and VDM polarization four-vectors,
 $a_{s}^\dagger (k)$  ($b_{r}^\dagger (q)$) and $a_{s}(k)$   ($b_{r}(q)$) are the creation and annihilation operators of photon (VDM), respectively. They satisfy the following commutation relations
\begin{equation}\label{aadag}
[a_{s}(p),a_{s'}^\dagger (p')]=(2\pi)^3 2p^0 \delta_{ss'}\delta^{(3)}({\bf{p}}-{\bf{p'}}),
\end{equation}
\begin{equation}\label{bbdag}
[b_{r}(q),b_{r'}^\dagger (q')]=(2\pi)^3 2q^0 \delta_{rr'}\delta^{(3)}({\bf{q}}-{\bf{q'}}).
\end{equation}
Using (\ref{A})-(\ref{D}) the interaction Hamiltonian can be cast into
\begin{eqnarray}
H^{0}_{I}(t)&=& H^{(1)}_{I}(t)+H^{(2)}_{I}(t)+H^{(3)}_{I}(t)\nonumber\\
&=&\frac{1}{2}\int\,{d\bf{q}}\,{d\bf{q'}}\,{d\bf{p}}\,{d\bf{p'}} (2\pi)^3  \delta^{(3)} ({\bf{q'}}+{\bf{p'}}-{\bf{q}}-{\bf{p}}) e^{i(q'^0 +p'^0 -q^0 - p^0)t} \nonumber\\
&\times& b_{r'}^\dagger(q') a_{s'}^\dagger(p') \bigg(g^2_{_{V}}\cos^2 \theta_{_{W}}\mathcal{M}_{1}
+2g_{_{V}}g'_{_{V}}\cos^2 \theta_{_{W}}\mathcal{M}_{2}\nonumber\\
&+&g'^2_{_{V}}\cos^2 \theta_{_{W}}\mathcal{M}_{3}\bigg)b_{r}(q) a_{s}(p),
\end{eqnarray}	
with 
\begin{eqnarray}
{d \bf{q}} \equiv \frac {d^3 \bf{q}}{(2\pi)^3\,2q^{0}},\,\,\,\,\,\,\,\,\,\, {d \bf{p}} \equiv \frac {d^3 \bf{p}}{(2\pi)^3\,2p^{0}},
\end{eqnarray}
for momentum of the VDM and photon, respectively, and
\begin{eqnarray}\label{m1}
\mathcal{M}_{1}&=&\frac{-g_{\nu\sigma}+(q+p)_{\nu}(q+p)_{\sigma}/m_{_{V'}}^{2}}{m_{_{V}}^2-m_{_{V'}}^2+2p.q}\nonumber\\
&\times&\bigg(p'^{\mu}p^{\rho}\varepsilon^{\nu}_{s'}(p')\varepsilon^{\sigma}_{s}(p)\varepsilon_{r'\mu}(q')\varepsilon_{r\rho}(q)
-p'^{\mu}p^{\sigma}\varepsilon^{\nu}_{s'}(p')\varepsilon^{\rho}_{s}(p)\varepsilon_{r'\mu}(q')\varepsilon_{r\rho}(q)
\nonumber\\
&-&p'^{\nu}p^{\rho}\varepsilon^{\mu}_{s'}(p')\varepsilon^{\sigma}_{s}(p)\varepsilon_{r'\mu}(q')\varepsilon_{r\rho}(q) +p'^{\nu}p^{\sigma}\varepsilon^{\mu}_{s'}(p')\varepsilon^{\rho}_{s}(p)\varepsilon_{r'\mu}(q')\varepsilon_{r\rho}(q)\bigg)\nonumber\\
&+&\frac{-g_{\nu\sigma}+(q-p')_{\nu}(q-p')_{\sigma}/m_{_{V'}}^{2}}{m_{_{V}}^2-m_{_{V'}}^2-2p'.q}\nonumber\\
&\times&\bigg(p^{\mu}p'^{\rho}\varepsilon^{\nu}_{s}(p)\varepsilon^{\sigma}_{s'}(p')\varepsilon_{r'\mu}(q')\varepsilon_{r\rho}(q)-p^{\mu}p'^{\sigma}\varepsilon^{\nu}_{s}(p)\varepsilon^{\rho}_{s'}(p')\varepsilon_{r'\mu}(q')\varepsilon_{r\rho}(q)\nonumber\\
&-&p^{\nu}p'^{\rho}\varepsilon^{\mu}_{s}(p)\varepsilon^{\sigma}_{s'}(p')\varepsilon_{r'\mu}(q')\varepsilon_{r\rho}(q) +p^{\nu}p'^{\sigma}\varepsilon^{\mu}_{s}(p)\varepsilon^{\rho}_{s'}(p')\varepsilon_{r'\mu}(q')\varepsilon_{r\rho}(q)\bigg),\nonumber\\
\end{eqnarray}
\begin{eqnarray}\label{m2}
\mathcal{M}_{2}&=& \varepsilon^{\rho\sigma\lambda\eta}\bigg\{\frac{-g_{\nu\eta}+(q+p)_{\nu}(q+p)_{\eta}/m_{_{V'}}^{2}}{m_{_{V}}^2-m_{_{V'}}^2+2p.q}\nonumber\\
&\times&\bigg((p.\varepsilon_{r}(q))p'_{\rho}\varepsilon_{s' \sigma}(p')\varepsilon^{\nu}_{s}(p) \varepsilon_{r' \lambda}(q')
-(\varepsilon_{s}(p).\varepsilon_{r}(q))p'_{\rho}p^{\nu}\varepsilon_{s'\sigma}(p')\varepsilon_{r'\lambda}(q')\nonumber\\
&+&(p'.\varepsilon_{r'}(q'))p_{\rho}\varepsilon^{\nu}_{s' }(p')\varepsilon_{s\sigma}(p) \varepsilon_{r \lambda}(q)
-(\varepsilon_{s'}(p').\varepsilon_{r'}(q'))p_{\rho}p'^{\nu}\varepsilon_{s\sigma}(p)\varepsilon_{r\lambda}(q)\bigg)\nonumber\\
&+&\frac{-g_{\nu\eta}+(q-p')_{\nu}(q-p')_{\eta}/m_{_{V'}}^{2}}{m_{_{V}}^2-m_{_{V'}}^2-2p'.q}\nonumber\\
&\times&\bigg((p'.\varepsilon_{r}(q))p_{\rho}\varepsilon_{s \sigma}(p)\varepsilon^{\nu}_{s'}(p') \varepsilon_{r' \lambda}(q')
-(\varepsilon_{s'}(p').\varepsilon_{r}(q))p_{\rho}p'^{\nu}\varepsilon_{s\sigma}(p)\varepsilon_{r'\lambda}(q')\nonumber\\
&+&(p.\varepsilon_{r'}(q'))p'_{\rho}\varepsilon^{\nu}_{s}(p)\varepsilon_{s'\sigma}(p') \varepsilon_{r \lambda}(q)
-(\varepsilon_{s}(p).\varepsilon_{r'}(q'))p'_{\rho}p^{\nu}\varepsilon_{s'\sigma}(p')\varepsilon_{r\lambda}(q)\bigg)\bigg\},\nonumber\\
\end{eqnarray}
\begin{eqnarray}\label{m3}
\mathcal{M}_{3}&=& 4\,\varepsilon^{\mu\upsilon\alpha\beta}\varepsilon^{\rho\sigma\lambda\eta}\nonumber\\
&\times&\bigg(\frac{-g_{\beta\eta}+(q+p)_{\eta}(q+p)_{\beta}/m_{_{V'}}^{2}}{m_{_{V}}^2-m_{_{V'}}^2+2p.q}\,p'_{\mu}\,p_{\rho}\varepsilon_{s' \nu}(p') \varepsilon_{s\sigma}(p)\varepsilon_{r' \alpha}(q')\varepsilon_{r\lambda}(q)\nonumber\\
&&+\frac{-g_{\beta\eta}+(q-p')_{\eta}(q-p')_{\beta}}{m_{_{V}}^2-m_{_{V'}}^2-2p'.q}\,p_{\mu}\,p'_{\rho}\varepsilon_{s \nu}(p) \varepsilon_{s'\sigma}(p')\varepsilon_{r' \alpha}(q')\varepsilon_{r\lambda}(q)\bigg).
\end{eqnarray}
By using (\ref{aadag}), the commutation relation between the interaction Hamiltonian  $H^0_{I}(t)$ and number operator $D_{ij}^0({\bf{k}})=a_i^\dagger({\bf{k}})a_j({\bf{k}})$ can be obtained as follows
\begin{eqnarray}\label{expecH&D}
[H^{0}_{I}(0),D^{0}_{ij}({\bf{k}})]&=&\frac{1}{2}\int \,{d\bf{q}}\, {d\bf{q'}}\,{d\bf{p}}\,{d\bf{p'}} (2\pi)^3 \delta^{(3)} ({\bf{q'}}+{\bf{p'}}-{\bf{q}}-{\bf{p}}) \nonumber\\
&\times&[b_{r'}^\dagger(q')b_{r}(q) a_{s'}^\dagger(p')a_{j}(k)2p^0(2\pi)^3\delta_{is}\delta^3(\bf{p}-\bf{k})\nonumber\\
&-& b_{r'}^\dagger(q')b_{r}(q) a_{i}^\dagger(k)a_{s}(p)2p'^0(2\pi)^3\delta_{js'}\delta^3(\bf{p'}-\bf{k})]\nonumber\\
&\times& \bigg(g^2_{_{V}}\cos^2 \theta_{_{W}}\mathcal{M}_{1}
+2g_{_{V}}g'_{_{V}}\cos^2 \theta_{_{W}}\mathcal{M}_{2}\nonumber\\
&+&g'^2_{_{V}}\cos^2 \theta_{_{W}}\mathcal{M}_{3}\bigg),
\end{eqnarray}
where by using 
\begin{eqnarray}
\langle a_{1}a_{2}...b_{1}b_{2}...\rangle&=& \langle a_{1}a_{2}...\rangle \langle b_{1}b_{2}...\rangle,\\
\langle a_{m}^\dagger(p')a_{n}(p)\rangle &=& 2p^0(2\pi)^3\delta^3({\bf{p}}-{\bf{p'}})\rho_{mn}({\bf{p}}),\\
\langle b_{m}^\dagger(q')b_{n}(q)\rangle &=& 2q^0(2\pi)^3\delta^3({\bf{q}}-{\bf{q'}}) \delta_{mn}\rho_{mn}({\bf{q}}),
\end{eqnarray}
one can obtain the expectation value of (\ref{expecH&D}) as
\begin{eqnarray}\label{forwardterm}
i\langle [H_{I}^0(0), D_{ij}({\bf{k}})]\rangle&=& \frac{i}{2}\int d{\bf{q}} \frac{2(m^2_{_{V}}-m^2_{_{V'}})}{{(m^2_{_{V}}-m^2_{_{V'}})^2-4(k.q)^2}}(2\pi)^3 \delta^3(0)\nonumber\\
&\times&\sum_{s,s'=1}^{2}\sum_{r=1}^{3}\bigg(\delta_{is}\rho_{s'j}({\bf{k}})-\delta_{s'j}\rho_{is}({\bf{k}})\bigg)\rho_{rr}({\bf{x}},{\bf{q}})\bigg[\frac{2g_{_{V}}g'_{_{V}}\cos^2 \theta_{_{W}}}{m^2_{_{V'}}}\varepsilon^{\rho\sigma\lambda\eta}\nonumber\\
& \times&\bigg( \left((k.\varepsilon_{r}(q))(q.\varepsilon_{s'}(k))-(k.q)(\varepsilon_{s'}(k).\varepsilon_{r}(q))\right)k_{\rho} \varepsilon_{s\sigma}(k) \varepsilon_{r\lambda}(q)q_{\eta}\nonumber\\
&+&\left((k.\varepsilon_{r}(q))(q.\varepsilon_{s}(k))-(k.q)(\varepsilon_{s}(k).\varepsilon_{r}(q))\right) k_{\rho}\varepsilon_{s'\sigma}(k) \varepsilon_{r\lambda}(q)q_{\eta}\bigg)\nonumber\\
&+&\frac{(g_{_{V}}^2-4g_{_{V}}^{'2})\cos^2 \theta_{_{W}}}{m^2_{_{V'}}}\bigg((k.\varepsilon_{r}(q))(q.\varepsilon_{s}(k))-(k.q)(\varepsilon_{s}(k).\varepsilon_{r}(q))\bigg)\nonumber\\
&\times&\bigg((k.\varepsilon_{r}(q))(q.\varepsilon_{s'}(k))-(k.q)(\varepsilon_{s'}(k).\varepsilon_{r}(q))\bigg)\nonumber\\
&+&\delta_{ss'} \bigg(\frac{4g_{_{V}}^{'2}\cos^2 \theta_{_{W}}}{m^2_{_{V'}}}((m^2_{_{V}}-m^2_{_{V'}})(k.\varepsilon_{r}(q))^2+(k.q)^2)\nonumber\\
&-&g_{_{V}}^2\cos^2 \theta_{_{W}}(k.\varepsilon_{r}(q))^2\bigg)\bigg].
\end{eqnarray}
Meanwhile, from  (\ref {roij}) the time evolution of the photon density matrix elements up to the leading term can be derived as follows
\begin{eqnarray}\label{roij3}
\frac{d\rho_{ij}}{dt}&=&\frac{i}{2k^{0}}\int d{\bf{q}}  \frac{(m^2_{_{V}}-m^2_{_{V'}})}{{(m^2_{_{V}}-m^2_{_{V'}})^2-4(k.q)^2}}\nonumber\\
&\times&\sum_{s,s'=1}^{2}\sum_{r=1}^{3}\bigg(\delta_{is}\rho_{s'j}({\bf{k}})-\delta_{s'j}\rho_{is}({\bf{k}})\bigg)\rho_{rr}({\bf{x}},{\bf{q}})\bigg[\frac{2g_{_{V}}g'_{_{V}}\cos^2 \theta_{_{W}}}{m^2_{_{V'}}}\varepsilon^{\rho\sigma\lambda\eta}\nonumber\nonumber\\
&\times&\bigg( \left((k.\varepsilon_{r}(q))(q.\varepsilon_{s'}(k))-(k.q)(\varepsilon_{s'}(k).\varepsilon_{r}(q))\right)k_{\rho} \varepsilon_{s\sigma}(k) \varepsilon_{r\lambda}(q)q_{\eta}\nonumber\\
&+&\left((k.\varepsilon_{r}(q))(q.\varepsilon_{s}(k))-(k.q)(\varepsilon_{s}(k).\varepsilon_{r}(q))\right) k_{\rho}\varepsilon_{s'\sigma}(k) \varepsilon_{r\lambda}(q)q_{\eta}\bigg)\nonumber\\
&+&\frac{(g_{_{V}}^2-4g_{_{V}}^{'2})\cos^2 \theta_{_{W}}}{m^2_{_{V'}}}\bigg((k.\varepsilon_{r}(q))(q.\varepsilon_{s}(k))-(k.q)(\varepsilon_{s}(k).\varepsilon_{r}(q))\bigg)\nonumber\\
&\times&\bigg((k.\varepsilon_{r}(q))(q.\varepsilon_{s'}(k))-(k.q)(\varepsilon_{s'}(k).\varepsilon_{r}(q))\bigg)+\delta_{ss'} \bigg(\frac{4g_{_{V}}^{'2}\cos^2 \theta_{_{W}}}{m^2_{_{V'}}}\nonumber\\
&\times&((m^2_{_{V}}-m^2_{_{V'}})(k.\varepsilon_{r}(q))^2+(k.q)^2)-g_{_{V}}^2\cos^2 \theta_{_{W}}(k.\varepsilon_{r}(q))^2\bigg)\bigg],
\end{eqnarray}
or in terms of its components one has
\begin{eqnarray}\label{ro11}
\frac{d\rho_{11}}{dt} &=&-\frac{1}{2k^{0}}\int d{\bf{q}}  \frac{(m^2_{_{V}}-m^2_{_{V'}})}{{(m^2_{_{V}}-m^2_{_{V'}})^2-4(k.q)^2}}\sum_{r=1}^{3}\rho_{rr}({\bf{x}},{\bf{q}})\bigg[\frac{2g_{_{V}}g'_{_{V}}\cos^2 \theta_{_{W}}}{m^2_{_{V'}}}\varepsilon^{\rho\sigma\lambda\eta}\nonumber\\
& \times&\bigg( \left((k.\varepsilon_{r}(q))(q.\varepsilon_{1}(k))-(k.q)(\varepsilon_{1}(k).\varepsilon_{r}(q))\right)k_{\rho} \varepsilon_{2\sigma}(k) \varepsilon_{r\lambda}(q)q_{\eta}\nonumber\\
&+&\left((k.\varepsilon_{r}(q))(q.\varepsilon_{2}(k))-(k.q)(\varepsilon_{2}(k).\varepsilon_{r}(q))\right) k_{\rho}\varepsilon_{1\sigma}(k) \varepsilon_{r\lambda}(q)q_{\eta}\bigg)\nonumber\\
&+&\frac{(g_{_{V}}^2-4g_{_{V}}^{'2})\cos^2 \theta_{_{W}}}{m^2_{_{V'}}}\bigg((k.\varepsilon_{r}(q))(q.\varepsilon_{1}(k))-(k.q)(\varepsilon_{1}(k).\varepsilon_{r}(q))\bigg)\nonumber\\
&\times&\bigg((k.\varepsilon_{r}(q))(q.\varepsilon_{2}(k))-(k.q)(\varepsilon_{2}(k).\varepsilon_{r}(q))\bigg)\bigg]V({\bf{k}}),
\end{eqnarray}

\begin{eqnarray}\label{ro22}
\frac{d\rho_{22}}{dt} &=&\frac{1}{2k^{0}}\int d{\bf{q}}  \frac{(m^2_{_{V}}-m^2_{_{V'}})}{{(m^2_{_{V}}-m^2_{_{V'}})^2-4(k.q)^2}}\sum_{r=1}^{3}\rho_{rr}({\bf{x}},{\bf{q}})\bigg[\frac{2g_{_{V}}g'_{_{V}}\cos^2 \theta_{_{W}}}{m^2_{_{V'}}}\varepsilon^{\rho\sigma\lambda\eta}\nonumber\\
& \times&\bigg( \left((k.\varepsilon_{r}(q))(q.\varepsilon_{1}(k))-(k.q)(\varepsilon_{1}(k).\varepsilon_{r}(q))\right)k_{\rho} \varepsilon_{2\sigma}(k) \varepsilon_{r\lambda}(q)q_{\eta}\nonumber\\
&+&\left((k.\varepsilon_{r}(q))(q.\varepsilon_{2}(k))-(k.q)(\varepsilon_{2}(k).\varepsilon_{r}(q))\right) k_{\rho}\varepsilon_{1\sigma}(k) \varepsilon_{r\lambda}(q)q_{\eta}\bigg)\nonumber\\
&+&\frac{(g_{_{V}}^2-4g_{_{V}}^{'2})\cos^2 \theta_{_{W}}}{m^2_{_{V'}}}\bigg((k.\varepsilon_{r}(q))(q.\varepsilon_{1}(k))-(k.q)(\varepsilon_{1}(k).\varepsilon_{r}(q))\bigg)\nonumber\\
&\times&\bigg((k.\varepsilon_{r}(q))(q.\varepsilon_{2}(k))-(k.q)(\varepsilon_{2}(k).\varepsilon_{r}(q))\bigg)\bigg]V({\bf{k}}),
\end{eqnarray}

\begin{eqnarray}\label{ro12}
\frac{d\rho_{12}}{dt} &=& \frac{i}{2k^{0}}\int d{\bf{q}}  \frac{(m^2_{_{V}}-m^2_{_{V'}})}{{(m^2_{_{V}}-m^2_{_{V'}})^2-4(k.q)^2}}\sum_{r=1}^{3}\rho_{rr}({\bf{x}},{\bf{q}})\bigg\{\bigg[\frac{4g_{_{V}}g'_{_{V}}\cos^2 \theta_{_{W}}}{m^2_{_{V'}}}\nonumber\\
& \times&\varepsilon^{\rho\sigma\lambda\eta}\nonumber\bigg( \left((k.\varepsilon_{r}(q))(q.\varepsilon_{1}(k))-(k.q)(\varepsilon_{1}(k).\varepsilon_{r}(q))\right)k_{\rho} \varepsilon_{1\sigma}(k) \varepsilon_{r\lambda}(q)q_{\eta}\nonumber\\
&-&\left((k.\varepsilon_{r}(q))(q.\varepsilon_{2}(k))-(k.q)(\varepsilon_{2}(k).\varepsilon_{r}(q))\right) k_{\rho}\varepsilon_{2\sigma}(k) \varepsilon_{r\lambda}(q)q_{\eta}\bigg)\nonumber\\
&+&\frac{(g_{_{V}}^2-4g_{_{V}}^{'2})\cos^2 \theta_{_{W}}}{m^2_{_{V'}}}\bigg(\bigg((k.\varepsilon_{r}(q))(q.\varepsilon_{1}(k))-(k.q)(\varepsilon_{1}(k).\varepsilon_{r}(q))\bigg)^2\nonumber\\
&-&\bigg((k.\varepsilon_{r}(q))(q.\varepsilon_{2}(k))-(k.q)(\varepsilon_{2}(k).\varepsilon_{r}(q))\bigg)^2\bigg)\bigg]\rho_{12}({\bf{k}})-\bigg[\frac{2g_{_{V}}g'_{_{V}}\cos^2 \theta_{_{W}}}{m^2_{_{V'}}}\nonumber\\
&\times&\varepsilon^{\rho\sigma\lambda\eta}\bigg( \left((k.\varepsilon_{r}(q))(q.\varepsilon_{1}(k))-(k.q)(\varepsilon_{1}(k).\varepsilon_{r}(q))\right)k_{\rho} \varepsilon_{2\sigma}(k) \varepsilon_{r\lambda}(q)q_{\eta}\nonumber\\
&+&\left((k.\varepsilon_{r}(q))(q.\varepsilon_{2}(k))-(k.q)(\varepsilon_{2}(k).\varepsilon_{r}(q))\right) k_{\rho}\varepsilon_{1\sigma}(k) \varepsilon_{r\lambda}(q)q_{\eta}\bigg)\nonumber\\
&+&\frac{(g_{_{V}}^2-4g_{_{V}}^{'2})\cos^2 \theta_{_{W}}}{m^2_{_{V'}}}\bigg((k.\varepsilon_{r}(q))(q.\varepsilon_{1}(k))-(k.q)(\varepsilon_{1}(k).\varepsilon_{r}(q))\bigg)\nonumber\\
&\times&\bigg((k.\varepsilon_{r}(q))(q.\varepsilon_{2}(k))-(k.q)(\varepsilon_{2}(k).\varepsilon_{r}(q))\bigg)\bigg]Q({\bf{k}}) \bigg\},
\end{eqnarray}

\begin{eqnarray}\label{ro21}
\frac{d\rho_{21}}{dt} &=& -\frac{i}{2k^{0}}\int d{\bf{q}}  \frac{(m^2_{_{V}}-m^2_{_{V'}})}{{(m^2_{_{V}}-m^2_{_{V'}})^2-4(k.q)^2}}\sum_{r=1}^{3}\rho_{rr}({\bf{x}},{\bf{q}})\bigg\{\bigg[\frac{4g_{_{V}}g'_{_{V}}\cos^2 \theta_{_{W}}}{m^2_{_{V'}}}\varepsilon^{\rho\sigma\lambda\eta}\nonumber\\
& \times&\bigg( \left((k.\varepsilon_{r}(q))(q.\varepsilon_{1}(k))-(k.q)(\varepsilon_{1}(k).\varepsilon_{r}(q))\right)k_{\rho} \varepsilon_{1\sigma}(k) \varepsilon_{r\lambda}(q)q_{\eta}\nonumber\\
&-&\left((k.\varepsilon_{r}(q))(q.\varepsilon_{2}(k))-(k.q)(\varepsilon_{2}(k).\varepsilon_{r}(q))\right) k_{\rho}\varepsilon_{2\sigma}(k) \varepsilon_{r\lambda}(q)q_{\eta}\bigg)\nonumber\\
&+&\frac{(g_{_{V}}^2-4g_{_{V}}^{'2})\cos^2 \theta_{_{W}}}{m^2_{_{V'}}}\bigg(\bigg((k.\varepsilon_{r}(q))(q.\varepsilon_{1}(k))-(k.q)(\varepsilon_{1}(k).\varepsilon_{r}(q))\bigg)^2\nonumber\\
&-&\bigg((k.\varepsilon_{r}(q))(q.\varepsilon_{2}(k))-(k.q)(\varepsilon_{2}(k).\varepsilon_{r}(q))\bigg)^2\bigg)\bigg]\rho_{21}({\bf{k}})-\bigg[\frac{2g_{_{V}}g'_{_{V}}\cos^2 \theta_{_{W}}}{m^2_{_{V'}}}\nonumber\\
& \times&\varepsilon^{\rho\sigma\lambda\eta}\nonumber\bigg( \left((k.\varepsilon_{r}(q))(q.\varepsilon_{1}(k))-(k.q)(\varepsilon_{1}(k).\varepsilon_{r}(q))\right)k_{\rho} \varepsilon_{2\sigma}(k) \varepsilon_{r\lambda}(q)q_{\eta}\nonumber\\
&+&\left((k.\varepsilon_{r}(q))(q.\varepsilon_{2}(k))-(k.q)(\varepsilon_{2}(k).\varepsilon_{r}(q))\right) k_{\rho}\varepsilon_{1\sigma}(k) \varepsilon_{r\lambda}(q)q_{\eta}\bigg)\nonumber\\
&+&\frac{(g_{_{V}}^2-4g_{_{V}}^{'2})\cos^2 \theta_{_{W}}}{m^2_{_{V'}}}\bigg((k.\varepsilon_{r}(q))(q.\varepsilon_{1}(k))-(k.q)(\varepsilon_{1}(k).\varepsilon_{r}(q))\bigg)\nonumber\\
&\times&\bigg((k.\varepsilon_{r}(q))(q.\varepsilon_{2}(k))-(k.q)(\varepsilon_{2}(k).\varepsilon_{r}(q))\bigg)\bigg]Q({\bf{k}}) \bigg\}.
\end{eqnarray}
Now, we are ready to determine the time evolution of Stocks parameters for the VDM-photon scattering as follows
\begin{equation}\label{I1}
\dot{I}({\bf{k}})=\dot{\rho}_{11}({\bf{k}})+\dot{\rho}_{22}({\bf{k}})=0,
\end{equation}

\begin{eqnarray}\label{Q1}
\dot{Q}({\bf{k}})&=&\dot{\rho}_{11}({\bf{k}})-\dot{\rho}_{22}({\bf{k}})\nonumber\\
&=&-\frac{1}{2k^{0}}\int d{\bf{q}} \frac{(m^2_{_{V}}-m^2_{_{V'}})}{{(m^2_{_{V}}-m^2_{_{V'}})^2-4(k.q)^2}}\sum_{r=1}^{3}\rho_{rr}({\bf{x}},{\bf{q}})\bigg[\frac{4g_{_{V}}g'_{_{V}}\cos^2 \theta_{_{W}}}{m^2_{_{V'}}}\varepsilon^{\rho\sigma\lambda\eta}\nonumber\\
&\times&\bigg( \left((k.\varepsilon_{r}(q))(q.\varepsilon_{1}(k))-(k.q)(\varepsilon_{1}(k).\varepsilon_{r}(q))\right)k_{\rho} \varepsilon_{2\sigma}(k) \varepsilon_{r\lambda}(q)q_{\eta}\nonumber\\
&+&\left((k.\varepsilon_{r}(q))(q.\varepsilon_{2}(k))-(k.q)(\varepsilon_{2}(k).\varepsilon_{r}(q))\right) k_{\rho}\varepsilon_{1\sigma}(k) \varepsilon_{r\lambda}(q)q_{\eta}\bigg)\nonumber\\
&+&\frac{2(g_{_{V}}^2-4g_{_{V}}^{'2})\cos^2 \theta_{_{W}}}{m^2_{_{V'}}}\bigg((k.\varepsilon_{r}(q))(q.\varepsilon_{1}(k))-(k.q)(\varepsilon_{1}(k).\varepsilon_{r}(q))\bigg)\nonumber\\
&\times&\bigg((k.\varepsilon_{r}(q))(q.\varepsilon_{2}(k))-(k.q)(\varepsilon_{2}(k).\varepsilon_{r}(q))\bigg)\bigg]V({\bf{k}}),
\end{eqnarray}

\begin{eqnarray}\label{U1}
\dot{U}({\bf{k}})&=&\dot{\rho}_{12}({\bf{k}})+\dot{\rho}_{21}({\bf{k}})\nonumber\\&=&
\frac{1}{2k^{0}}\int d{\bf{q}} \frac{(m^2_{_{V}}-m^2_{_{V'}})}{{(m^2_{_{V}}-m^2_{_{V'}})^2-4(k.q)^2}}\sum_{r=1}^{3}\rho_{rr}({\bf{x}},{\bf{q}})\bigg[\frac{4g_{_{V}}g'_{_{V}}\cos^2 \theta_{_{W}}}{m^2_{_{V'}}}\varepsilon^{\rho\sigma\lambda\eta}\nonumber\\
& \times&\bigg( \left((k.\varepsilon_{r}(q))(q.\varepsilon_{1}(k))-(k.q)(\varepsilon_{1}(k).\varepsilon_{r}(q))\right)k_{\rho} \varepsilon_{1\sigma}(k) \varepsilon_{r\lambda}(q)q_{\eta}\nonumber\\
&-&\left((k.\varepsilon_{r}(q))(q.\varepsilon_{2}(k))-(k.q)(\varepsilon_{2}(k).\varepsilon_{r}(q))\right) k_{\rho}\varepsilon_{2\sigma}(k) \varepsilon_{r\lambda}(q)q_{\eta}\bigg)\nonumber\\
&+&\frac{(g_{_{V}}^2-4g_{_{V}}^{'2})\cos^2 \theta_{_{W}}}{m^2_{_{V'}}}\bigg(\bigg((k.\varepsilon_{r}(q))(q.\varepsilon_{1}(k))-(k.q)(\varepsilon_{1}(k).\varepsilon_{r}(q))\bigg)^2\nonumber\\
&-&\bigg((k.\varepsilon_{r}(q))(q.\varepsilon_{2}(k))-(k.q)(\varepsilon_{2}(k).\varepsilon_{r}(q))\bigg)^2\bigg)\bigg]V({\bf{k}}),
\end{eqnarray}
and
\begin{eqnarray} \label{V1}
\dot{V}({\bf{k}})&=& i\dot{\rho}_{12}({\bf{k}})-i\dot{\rho}_{21}({\bf{k}})\nonumber\\
&=&-\frac{1}{2k^{0}}\int d{\bf{q}} \frac{(m^2_{_{V}}-m^2_{_{V'}})}{{(m^2_{_{V}}-m^2_{_{V'}})^2-4(k.q)^2}}\sum_{r=1}^{3}\rho_{rr}({\bf{x}},{\bf{q}})\bigg\{\bigg[\frac{4g_{_{V}}g'_{_{V}}\cos^2 \theta_{_{W}}}{m^2_{_{V'}}}\varepsilon^{\rho\sigma\lambda\eta}\nonumber\\
&\times&\bigg( \left((k.\varepsilon_{r}(q))(q.\varepsilon_{1}(k))-(k.q)(\varepsilon_{1}(k).\varepsilon_{r}(q))\right)k_{\rho} \varepsilon_{1\sigma}(k) \varepsilon_{r\lambda}(q)q_{\eta}\nonumber\\
&-&\left((k.\varepsilon_{r}(q))(q.\varepsilon_{2}(k))-(k.q)(\varepsilon_{2}(k).\varepsilon_{r}(q))\right) k_{\rho}\varepsilon_{2\sigma}(k) \varepsilon_{r\lambda}(q)q_{\eta}\bigg)\nonumber\\
&+&\frac{(g_{_{V}}^2-4g_{_{V}}^{'2})\cos^2 \theta_{_{W}}}{m^2_{_{V'}}}\bigg(\bigg((k.\varepsilon_{r}(q))(q.\varepsilon_{1}(k))-(k.q)(\varepsilon_{1}(k).\varepsilon_{r}(q))\bigg)^2\nonumber\\
&-&\bigg((k.\varepsilon_{r}(q))(q.\varepsilon_{2}(k))-(k.q)(\varepsilon_{2}(k).\varepsilon_{r}(q))\bigg)^2\bigg)\bigg]U({\bf{k}})-\bigg[\frac{4g_{_{V}}g'_{_{V}}\cos^2 \theta_{_{W}}}{m^2_{_{V'}}}\varepsilon^{\rho\sigma\lambda\eta}\nonumber\\
&\times&\bigg( \left((k.\varepsilon_{r}(q))(q.\varepsilon_{1}(k))-(k.q)(\varepsilon_{1}(k).\varepsilon_{r}(q))\right)k_{\rho} \varepsilon_{2\sigma}(k) \varepsilon_{r\lambda}(q)q_{\eta}\nonumber\\
&+&\left((k.\varepsilon_{r}(q))(q.\varepsilon_{2}(k))-(k.q)(\varepsilon_{2}(k).\varepsilon_{r}(q))\right) k_{\rho}\varepsilon_{1\sigma}(k) \varepsilon_{r\lambda}(q)q_{\eta}\bigg)\nonumber\\
&+&\frac{2(g_{_{V}}^2-4g_{_{V}}^{'2})\cos^2 \theta_{_{W}}}{m^2_{_{V'}}}\bigg((k.\varepsilon_{r}(q))(q.\varepsilon_{1}(k))-(k.q)(\varepsilon_{1}(k).\varepsilon_{r}(q))\bigg)\nonumber\\
&\times&\bigg((k.\varepsilon_{r}(q))(q.\varepsilon_{2}(k))-(k.q)(\varepsilon_{2}(k).\varepsilon_{r}(q))\bigg)\bigg]Q({\bf{k}}) \bigg\}.
\end{eqnarray}
To perform the integrals in  (\ref{Q1})-(\ref{V1}),  we
assume a uniform distribution function $\rho^{0}({\bf{x}},{\bf{q}})$ for the VDM such that the density matrix elements  $\rho_{rr}({\bf{x}},{\bf{q}})=\frac{1}{3}\rho^{0}({\bf{x}},{\bf{q}}),\,\,\, r=1,2,3$ and consequently the time evolution for the Stocks parameters can be obtained as
\begin{equation}\label{I2}
\dot{I}({\bf{k}})=0
\end{equation}
\begin{eqnarray}\label{Q2}
\dot{Q}({\bf{k}})&=&-\frac{1}{6k^{0}}\int d{\bf{q}}  \frac{(m^2_{_{V}}-m^2_{_{V'}})}{{(m^2_{_{V}}-m^2_{_{V'}})^2-4(k.q)^2}}\rho^{0}({\bf{x}},{\bf{q}})\bigg[\frac{4g_{_{V}}g'_{_{V}}\cos^2 \theta_{_{W}}}{m^2_{_{V'}}}\nonumber\\
& \times&\varepsilon^{\rho\sigma\lambda\eta}\bigg( \left((k.\varepsilon_{1}(q))(q.\varepsilon_{1}(k))-(k.q)(\varepsilon_{1}(k).\varepsilon_{1}(q))\right)k_{\rho} \varepsilon_{2\sigma}(k) \varepsilon_{1\lambda}(q)q_{\eta}\nonumber\\
&+&\left((k.\varepsilon_{1}(q))(q.\varepsilon_{2}(k))-(k.q)(\varepsilon_{2}(k).\varepsilon_{1}(q))\right) k_{\rho}\varepsilon_{1\sigma}(k) \varepsilon_{1\lambda}(q)q_{\eta}\nonumber\\
&+& \left((k.\varepsilon_{2}(q))(q.\varepsilon_{1}(k))-(k.q)(\varepsilon_{1}(k).\varepsilon_{2}(q))\right)k_{\rho} \varepsilon_{2\sigma}(k) \varepsilon_{2\lambda}(q)q_{\eta}\nonumber\\
&+&\left((k.\varepsilon_{2}(q))(q.\varepsilon_{2}(k))-(k.q)(\varepsilon_{2}(k).\varepsilon_{2}(q))\right) k_{\rho}\varepsilon_{1\sigma}(k) \varepsilon_{2\lambda}(q)q_{\eta}\nonumber\\
&+& \left((k.\varepsilon_{3}(q))(q.\varepsilon_{1}(k))-(k.q)(\varepsilon_{1}(k).\varepsilon_{3}(q))\right)k_{\rho} \varepsilon_{2\sigma}(k) \varepsilon_{3\lambda}(q)q_{\eta}\nonumber\\
&+&\left((k.\varepsilon_{3}(q))(q.\varepsilon_{2}(k))-(k.q)(\varepsilon_{2}(k).\varepsilon_{3}(q))\right) k_{\rho}\varepsilon_{1\sigma}(k) \varepsilon_{3\lambda}(q)q_{\eta}\bigg)\nonumber\\
&+&\frac{2(g_{_{V}}^2-4g_{_{V}}^{'2})\cos^2 \theta_{_{W}}}{m^2_{_{V'}}}\bigg(\bigg((k.\varepsilon_{1}(q))(q.\varepsilon_{1}(k))-(k.q)(\varepsilon_{1}(k).\varepsilon_{1}(q))\bigg)\nonumber\\
&\times&\bigg((k.\varepsilon_{1}(q))(q.\varepsilon_{2}(k))-(k.q)(\varepsilon_{2}(k).\varepsilon_{1}(q))\bigg)\nonumber\\
&+&\bigg((k.\varepsilon_{2}(q))(q.\varepsilon_{1}(k))-(k.q)(\varepsilon_{1}(k).\varepsilon_{2}(q))\bigg)\nonumber\\
&\times&\bigg((k.\varepsilon_{2}(q))(q.\varepsilon_{2}(k))-(k.q)(\varepsilon_{2}(k).\varepsilon_{2}(q))\bigg)\nonumber\\
&+&\bigg((k.\varepsilon_{3}(q))(q.\varepsilon_{1}(k))-(k.q)(\varepsilon_{1}(k).\varepsilon_{3}(q))\bigg)\nonumber\\
&\times&\bigg((k.\varepsilon_{3}(q))(q.\varepsilon_{2}(k))-(k.q)(\varepsilon_{2}(k).\varepsilon_{3}(q))\bigg)\bigg)\bigg]V({\bf{k}}),
\end{eqnarray}

\begin{eqnarray}\label{U2}
\dot{U}({\bf{k}})&=&\frac{1}{6k^{0}}\int d{\bf{q}}  \frac{(m^2_{_{V}}-m^2_{_{V'}})}{{(m^2_{_{V}}-m^2_{_{V'}})^2-4(k.q)^2}}\rho^{0}({\bf{x}},{\bf{q}})\bigg[\frac{4g_{_{V}}g'_{_{V}}\cos^2 \theta_{_{W}}}{m^2_{_{V'}}}\nonumber\\
&\times&\varepsilon^{\rho\sigma\lambda\eta}\bigg( \left((k.\varepsilon_{1}(q))(q.\varepsilon_{1}(k))-(k.q)(\varepsilon_{1}(k).\varepsilon_{1}(q))\right)k_{\rho} \varepsilon_{1\sigma}(k) \varepsilon_{1\lambda}(q)q_{\eta}\nonumber\\
&-&\left((k.\varepsilon_{1}(q))(q.\varepsilon_{2}(k))-(k.q)(\varepsilon_{2}(k).\varepsilon_{1}(q))\right) k_{\rho}\varepsilon_{2\sigma}(k) \varepsilon_{1\lambda}(q)q_{\eta}\nonumber\\
&+& \left((k.\varepsilon_{2}(q))(q.\varepsilon_{1}(k))-(k.q)(\varepsilon_{1}(k).\varepsilon_{2}(q))\right)k_{\rho} \varepsilon_{1\sigma}(k) \varepsilon_{2\lambda}(q)q_{\eta}\nonumber\\
&-&\left((k.\varepsilon_{2}(q))(q.\varepsilon_{2}(k))-(k.q)(\varepsilon_{2}(k).\varepsilon_{2}(q))\right) k_{\rho}\varepsilon_{2\sigma}(k) \varepsilon_{2\lambda}(q)q_{\eta}\nonumber\\
&+& \left((k.\varepsilon_{3}(q))(q.\varepsilon_{1}(k))-(k.q)(\varepsilon_{1}(k).\varepsilon_{3}(q))\right)k_{\rho} \varepsilon_{1\sigma}(k) \varepsilon_{3\lambda}(q)q_{\eta}\nonumber\\
&-&\left((k.\varepsilon_{3}(q))(q.\varepsilon_{2}(k))-(k.q)(\varepsilon_{2}(k).\varepsilon_{3}(q))\right) k_{\rho}\varepsilon_{2\sigma}(k) \varepsilon_{3\lambda}(q)q_{\eta}\bigg)\nonumber\\
&+&\frac{(g_{_{V}}^2-4g_{_{V}}^{'2})\cos^2 \theta_{_{W}}}{m^2_{_{V'}}}\bigg(\bigg((k.\varepsilon_{1}(q))(q.\varepsilon_{1}(k))-(k.q)(\varepsilon_{1}(k).\varepsilon_{1}(q))\bigg)^2\nonumber\\
&-&\bigg((k.\varepsilon_{1}(q))(q.\varepsilon_{2}(k))-(k.q)(\varepsilon_{2}(k).\varepsilon_{r}(q))\bigg)^2\nonumber\\
&+&\bigg((k.\varepsilon_{2}(q))(q.\varepsilon_{1}(k))-(k.q)(\varepsilon_{1}(k).\varepsilon_{2}(q))\bigg)^2\nonumber\\
&-&\bigg((k.\varepsilon_{2}(q))(q.\varepsilon_{2}(k))-(k.q)(\varepsilon_{2}(k).\varepsilon_{2}(q))\bigg)^2\nonumber\\
&+&\bigg((k.\varepsilon_{3}(q))(q.\varepsilon_{1}(k))-(k.q)(\varepsilon_{1}(k).\varepsilon_{3}(q))\bigg)^2\nonumber\\
&-&\bigg((k.\varepsilon_{3}(q))(q.\varepsilon_{2}(k))-(k.q)(\varepsilon_{2}(k).\varepsilon_{3}(q))\bigg)^2
\bigg)\bigg]V({\bf{k}}),
 \end{eqnarray}

\begin{eqnarray} \label{V2}
\dot{V}({\bf{k}})&=& -\frac{1}{6k^{0}}\int d{\bf{q}}  \frac{(m^2_{_{V}}-m^2_{_{V'}})}{{(m^2_{_{V}}-m^2_{_{V'}})^2-4(k.q)^2}}\rho^{0}({\bf{x}},{\bf{q}})\bigg\{\bigg[\frac{4g_{_{V}}g'_{_{V}}\cos^2 \theta_{_{W}}}{m^2_{_{V'}}}\nonumber\\
& \times&\varepsilon^{\rho\sigma\lambda\eta}\bigg( \left((k.\varepsilon_{1}(q))(q.\varepsilon_{1}(k))-(k.q)(\varepsilon_{1}(k).\varepsilon_{1}(q))\right)k_{\rho} \varepsilon_{1\sigma}(k) \varepsilon_{1\lambda}(q)q_{\eta}\nonumber\\
&-&\left((k.\varepsilon_{1}(q))(q.\varepsilon_{2}(k))-(k.q)(\varepsilon_{2}(k).\varepsilon_{1}(q))\right) k_{\rho}\varepsilon_{2\sigma}(k) \varepsilon_{1\lambda}(q)q_{\eta}\nonumber\\
&+& \left((k.\varepsilon_{2}(q))(q.\varepsilon_{1}(k))-(k.q)(\varepsilon_{1}(k).\varepsilon_{2}(q))\right)k_{\rho} \varepsilon_{1\sigma}(k) \varepsilon_{2\lambda}(q)q_{\eta}\nonumber\\
&-&\left((k.\varepsilon_{2}(q))(q.\varepsilon_{2}(k))-(k.q)(\varepsilon_{2}(k).\varepsilon_{2}(q))\right) k_{\rho}\varepsilon_{2\sigma}(k) \varepsilon_{2\lambda}(q)q_{\eta}\nonumber\\
&+& \left((k.\varepsilon_{3}(q))(q.\varepsilon_{1}(k))-(k.q)(\varepsilon_{1}(k).\varepsilon_{3}(q))\right)k_{\rho} \varepsilon_{1\sigma}(k) \varepsilon_{3\lambda}(q)q_{\eta}\nonumber\\
&-&\left((k.\varepsilon_{3}(q))(q.\varepsilon_{2}(k))-(k.q)(\varepsilon_{2}(k).\varepsilon_{3}(q))\right) k_{\rho}\varepsilon_{2\sigma}(k) \varepsilon_{3\lambda}(q)q_{\eta}\bigg)\nonumber\\
&+&\frac{(g_{_{V}}^2-4g_{_{V}}^{'2})\cos^2 \theta_{_{W}}}{m^2_{_{V'}}}\bigg(\bigg((k.\varepsilon_{1}(q))(q.\varepsilon_{1}(k))-(k.q)(\varepsilon_{1}(k).\varepsilon_{1}(q))\bigg)^2\nonumber\\
&-&\bigg((k.\varepsilon_{1}(q))(q.\varepsilon_{2}(k))-(k.q)(\varepsilon_{2}(k).\varepsilon_{r}(q))\bigg)^2\nonumber\\
&+&\bigg((k.\varepsilon_{2}(q))(q.\varepsilon_{1}(k))-(k.q)(\varepsilon_{1}(k).\varepsilon_{2}(q))\bigg)^2\nonumber\\
&-&\bigg((k.\varepsilon_{2}(q))(q.\varepsilon_{2}(k))-(k.q)(\varepsilon_{2}(k).\varepsilon_{2}(q))\bigg)^2\nonumber\\
&+&\bigg((k.\varepsilon_{3}(q))(q.\varepsilon_{1}(k))-(k.q)(\varepsilon_{1}(k).\varepsilon_{3}(q))\bigg)^2\nonumber\\
&-&\bigg((k.\varepsilon_{3}(q))(q.\varepsilon_{2}(k))-(k.q)(\varepsilon_{2}(k).\varepsilon_{3}(q))\bigg)^2
\bigg)\bigg]U({\bf{k}})-\bigg[\frac{4g_{_{V}}g'_{_{V}}\cos^2 \theta_{_{W}}}{m^2_{_{V'}}}\nonumber\\
&\times&\varepsilon^{\rho\sigma\lambda\eta}\bigg( \left((k.\varepsilon_{1}(q))(q.\varepsilon_{1}(k))-(k.q)(\varepsilon_{1}(k).\varepsilon_{1}(q))\right)k_{\rho} \varepsilon_{2\sigma}(k) \varepsilon_{1\lambda}(q)q_{\eta}\nonumber\\
&+&\left((k.\varepsilon_{1}(q))(q.\varepsilon_{2}(k))-(k.q)(\varepsilon_{2}(k).\varepsilon_{1}(q))\right) k_{\rho}\varepsilon_{1\sigma}(k) \varepsilon_{1\lambda}(q)q_{\eta}\nonumber\\
&+& \left((k.\varepsilon_{2}(q))(q.\varepsilon_{1}(k))-(k.q)(\varepsilon_{1}(k).\varepsilon_{2}(q))\right)k_{\rho} \varepsilon_{2\sigma}(k) \varepsilon_{2\lambda}(q)q_{\eta}\nonumber\\
&+&\left((k.\varepsilon_{2}(q))(q.\varepsilon_{2}(k))-(k.q)(\varepsilon_{2}(k).\varepsilon_{2}(q))\right) k_{\rho}\varepsilon_{1\sigma}(k) \varepsilon_{2\lambda}(q)q_{\eta}\nonumber\\
&+& \left((k.\varepsilon_{3}(q))(q.\varepsilon_{1}(k))-(k.q)(\varepsilon_{1}(k).\varepsilon_{3}(q))\right)k_{\rho} \varepsilon_{2\sigma}(k) \varepsilon_{3\lambda}(q)q_{\eta}\nonumber\\
&+&\left((k.\varepsilon_{3}(q))(q.\varepsilon_{2}(k))-(k.q)(\varepsilon_{2}(k).\varepsilon_{3}(q))\right) k_{\rho}\varepsilon_{1\sigma}(k) \varepsilon_{3\lambda}(q)q_{\eta}\bigg)\nonumber\\
&+&\frac{2(g_{_{V}}^2-4g_{_{V}}^{'2})\cos^2 \theta_{_{W}}}{m^2_{_{V'}}}\bigg(\bigg((k.\varepsilon_{1}(q))(q.\varepsilon_{1}(k))-(k.q)(\varepsilon_{1}(k).\varepsilon_{1}(q))\bigg)\nonumber\\
&\times&\bigg((k.\varepsilon_{1}(q))(q.\varepsilon_{2}(k))-(k.q)(\varepsilon_{2}(k).\varepsilon_{1}(q))\bigg)\nonumber\\
&+&\bigg((k.\varepsilon_{2}(q))(q.\varepsilon_{1}(k))-(k.q)(\varepsilon_{1}(k).\varepsilon_{2}(q))\bigg)\nonumber\\
&\times&\bigg((k.\varepsilon_{2}(q))(q.\varepsilon_{2}(k))-(k.q)(\varepsilon_{2}(k).\varepsilon_{2}(q))\bigg)\nonumber\\
&+&\bigg((k.\varepsilon_{3}(q))(q.\varepsilon_{1}(k))-(k.q)(\varepsilon_{1}(k).\varepsilon_{3}(q))\bigg)\nonumber\\
&\times&\bigg((k.\varepsilon_{3}(q))(q.\varepsilon_{2}(k))-(k.q)(\varepsilon_{2}(k).\varepsilon_{3}(q))\bigg)\bigg)\bigg]Q({\bf{k}}) \bigg\}.
\end{eqnarray}
If we consider general choices for the momentum direction and polarization vectors of the DM and incoming photon which is equal to the outgoing photon in case of the forward scattering as 
\begin{eqnarray}\label{Eq1}
q&=&(q^0,| {\bf{q}}|\sin\theta\cos\phi,| {\bf{q}}|\sin\theta\sin\phi,| {\bf{q}}|\cos\theta),\\
\varepsilon_{1}(q)&=&(0, \cos\theta\cos\phi,\cos\theta\sin\phi,-\sin\theta),\\
\varepsilon_{2}(q)&=&(0, -\sin\phi,\cos\phi,0),\\
\varepsilon_{3}(q)&=&\frac{1}{m_{_{V}}}(|{\bf q}|, q^0\sin\theta\cos\phi,q^0\sin\theta\sin\phi,q^0\cos\theta),
\end{eqnarray}
\begin{eqnarray}\label{Eq2}
k&=&(k^0,| {\bf{k}}|\sin\alpha\cos\beta,| {\bf{k}}|\sin\alpha\sin\beta,| {\bf{k}}|\cos\alpha),\\
\varepsilon_{1}(k)&=&(0, \cos\alpha\cos\beta,\cos\alpha\sin\beta,-\sin\alpha),\\
\varepsilon_{2}(k)&=&(0, -\sin\beta,\cos\beta,0),
\end{eqnarray}
then the time evolution of the Stockes parameters lead to
\begin{eqnarray} \label{stoke111}
\dot{I}({\bf{k}})=0,
\end{eqnarray}
\begin{eqnarray} \label{stoke112}
\dot{Q}({\bf{k}})=0,
\end{eqnarray}
\begin{eqnarray} \label{stoke113}
\dot{U}({\bf{k}})=0,
\end{eqnarray}
\begin{eqnarray} \label{stoke114}
\dot{V}({\bf{k}})=0.
\end{eqnarray}
The relations (\ref{stoke111})-(\ref{stoke114}) show that the photon polarization can not be generated by the forward scattering of photon from the unpolarized VDM.

If we consider  a polarized VDM with density matrix as follows
\begin{equation}\label{matrisro}
\rho_{rr}(\bf{x},\bf{q}) =
\bigg\lbrace 
\begin{array}{rr}
(\frac{1}{3}+{\delta}_{r}) \,\,\rho^{0}({\bf{x}},{\bf{q}})\,& r=r' \\\\
0\,\,\,\,\,\,\,\,\,\,\,\,\,\,\,\,\,\,\,\, &  r\neq r' 
\end{array}
\end{equation}
which satisfies $\sum_{r}^{3}{\delta}_{r}=0$, the time evolution of the Stocks parameters for the lowest order contribution of multipole expansion of the angular function in terms of Legendre polynomials $l=0$ can be obtained as follows
\begin{eqnarray} \label{I4}
\dot{I}({\bf{k}})&=& 0
\end{eqnarray}
\begin{eqnarray} \label{q4}
\dot{Q}({\bf{k}})&=&-\frac{2\pi g_{_{V}}g_{_{V'}}\cos^2\theta_{_{W}}}{m_{_{V'}}^2}k^{0}\sin^2\alpha\,(\delta_{1}-\delta_{2})V({\bf{k}})\nonumber\\
&\times&\int d{\bf{q}} \frac{(m^2_{_{V}}-m^2_{_{V'}})}{{(m^2_{_{V}}-m^2_{_{V'}})^2-4(k.q)^2}}(q^{0^{2}}+| {\bf{q}}|^2)
\end{eqnarray}
\begin{eqnarray} \label{u4}
\dot{U}({\bf{k}})&=&\frac{\pi (g_{_{V}}^2-4g_{_{V'}}^2)\cos^2\theta_{_{W}}}{2m_{_{V'}}^2}k^{0}\sin^2\alpha\,(\delta_{1}-\delta_{2})V({\bf{k}})\nonumber\\
&\times&\int d{\bf{q}} \frac{(m^2_{_{V}}-m^2_{_{V'}})}{{(m^2_{_{V}}-m^2_{_{V'}})^2-4(k.q)^2}}(q^{0^{2}}+| {\bf{q}}|^2)
\end{eqnarray}
\begin{eqnarray}\label{V4}
\dot{V}({\bf{k}})&=&-\frac{\pi}{2}k^{0}\sin^2\alpha\,(\delta_{1}-\delta_{2})\bigg(\frac{4g_{_{V}}g_{_{V'}}}{m^2_{_{V'}}}Q({\bf{k}})-\frac{(g_{_{V}}^2 -
	4 g_{_{V'}}^2)}{m^2_{_{V'}}}U({\bf{k}})\bigg)\cos^2\theta_{_{W}}\nonumber\\
&\times&\int d{\bf{q}} \frac{(m^2_{_{V}}-m^2_{_{V'}})}{{(m^2_{_{V}}-m^2_{_{V'}})^2-4(k.q)^2}}(q^{0^{2}}+| {\bf{q}}|^2)
\end{eqnarray}
The relations (\ref{I4})-(\ref{V4}) in the  non-relativistic limits  $q^0=m_{_{V}}$ and $(m^2_{_{V}}-m^2_{_{V'}})^2\gg4(k.q)^2$ can be simplified as
\begin{eqnarray} \label{I120}
\dot{I}({\bf{k}})&=& 0\\
\label{Q120}
\dot{Q}({\bf{k}})&=& 2\pi g_{_{V}}g_{_{V'}}\cos^2\theta_{_{W}}\frac{ k^{0}m_{_{V}}\sin^2\alpha}{m_{_{V'}}^2(m^2_{_{V'}}-m^2_{_{V}})}\,(\delta_{1}-\delta_{2})V({\bf{k}})\nonumber\\
&\times&\int \frac{d^3{\bf q}}{(2\pi)^3}
\rho^0({\bf{x}},{\bf{q}}), 
\\ \label{U120}
\dot{U}({\bf{k}})&=&-\frac{\pi}{2}(g_{_{V}}^2 -
4 g_{_{V'}}^2)\cos^2\theta_{_{W}}\frac{ k^{0}m_{_{V}}\sin^2\alpha}{m_{_{V'}}^2(m^2_{_{V'}}-m^2_{_{V}})}(\delta_{1}-\delta_{2})V({\bf{k}})\nonumber\\
&\times&\int \frac{d^3{\bf q}}{(2\pi)^3}
\rho^0({\bf{x}},{\bf{q}}), 
\\ \label{V120}
\dot{V}({\bf{k}})&=&-\frac{\pi}{2}\frac{ k^{0}m_{_{V}}\sin^2\alpha}{m_{_{V'}}^2(m^2_{_{V'}}-m^2_{_{V}})}(\delta_{1}-\delta_{2})\bigg(4g_{_{V}}g_{_{V'}}Q({\bf{k}})-(g_{_{V}}^2 -
4 g_{_{V'}}^2)U({\bf{k}})\bigg)\cos^2\theta_{_{W}}\nonumber\\
&\times&\int \frac{d^3{\bf q}}{(2\pi)^3}
\rho^0({\bf{x}},{\bf{q}}).
\end{eqnarray}
Also, one can be obtained the time evolution of the Stocks parameters for next non-zero contribution $l=2$ as follows
\begin{eqnarray} \label{I5}
\dot{I}({\bf{k}})&=& 0
\end{eqnarray}
\begin{eqnarray} \label{q5}
\dot{Q}({\bf{k}})&=&\frac{4\pi g_{_{V}}g_{_{V'}}\cos^2\theta_{_{W}}}{5m_{_{V'}}^2}k^{0}\sin^2\alpha V({\bf{k}})\nonumber\\
&\times&\int d{\bf{q}} \frac{(m^2_{_{V}}-m^2_{_{V'}})}{{(m^2_{_{V}}-m^2_{_{V'}})^2-4(k.q)^2}}(\delta_{1}q^{0^{2}}-\delta_{2}| {\bf{q}}|^2-\delta_{3}m_{_{V}}^2)P_{2}(\cos\theta),
\end{eqnarray}
\begin{eqnarray} \label{u5}
\dot{U}({\bf{k}})&=&-\frac{\pi (g_{_{V}}^2-4g_{_{V'}}^2)\cos^2\theta_{_{W}}}{5m_{_{V'}}^2}k^{0}\sin^2\alpha V({\bf{k}})\nonumber\\
&\times&\int d{\bf{q}} \frac{(m^2_{_{V}}-m^2_{_{V'}})}{{(m^2_{_{V}}-m^2_{_{V'}})^2-4(k.q)^2}}(\delta_{1}q^{0^{2}}-\delta_{2}| {\bf{q}}|^2-\delta_{3}m_{_{V}}^2)P_{2}(\cos\theta),
\end{eqnarray}
\begin{eqnarray}\label{V5}
\dot{V}({\bf{k}})&=&-\frac{\pi}{5}k^{0}\sin^2\alpha\,\bigg(\frac{4g_{_{V}}g_{_{V'}}}{m^2_{_{V'}}}Q({\bf{k}})-\frac{(g_{_{V}}^2 -
	4 g_{_{V'}}^2)}{m^2_{_{V'}}}U({\bf{k}})\bigg)\cos^2\theta_{_{W}}\nonumber\\
&\times&\int d{\bf{q}} \frac{(m^2_{_{V}}-m^2_{_{V'}})}{{(m^2_{_{V}}-m^2_{_{V'}})^2-4(k.q)^2}}(\delta_{1}q^{0^{2}}-\delta_{2}| {\bf{q}}|^2-\delta_{3}m_{_{V}}^2)P_{2}(\cos\theta).
\end{eqnarray}
As the previous case, (\ref{I5})-(\ref{V5}) in the  non-relativistic limits  $q^0=m_{_{V}}$ and $(m^2_{_{V}}-m^2_{_{V'}})^2\gg4(k.q)^2$ cast into
\begin{eqnarray} \label{I121}
\dot{I}({\bf{k}})&=& 0\\
\label{Q121}
\dot{Q}({\bf{k}})&=&-\frac{4\pi}{5} g_{_{V}}g_{_{V'}}\cos^2\theta_{_{W}}\frac{ k^{0}m_{_{V}}\sin^2\alpha}{m_{_{V'}}^2(m^2_{_{V'}}-m^2_{_{V}})}\,(\delta_{1}-\delta_{3})V({\bf{k}})\nonumber\\
&\times&\int \frac{d^3{\bf q}}{(2\pi)^3}
\rho^0({\bf{x}},{\bf{q}})P_{2}(\cos\theta)
\\ \label{U121}
\dot{U}({\bf{k}})&=&\frac{\pi}{5} (g_{_{V}}^2-4g_{_{V'}}^2)\cos^2\theta_{_{W}}\frac{ k^{0}m_{_{V}}\sin^2\alpha}{m_{_{V'}}^2(m^2_{_{V'}}-m^2_{_{V}})}\,(\delta_{1}-\delta_{3})V({\bf{k}})\nonumber\\
&\times&\int \frac{d^3{\bf q}}{(2\pi)^3}
\rho^0({\bf{x}},{\bf{q}})P_{2}(\cos\theta),
\\ \label{V121}
\dot{V}({\bf{k}})&=&\frac{\pi}{5}\frac{ k^{0}m_{_{V}}\sin^2\alpha}{m_{_{V'}}^2(m^2_{_{V'}}-m^2_{_{V}})}(\delta_{1}-\delta_{3})\bigg(4g_{_{V}}g_{_{V'}}Q({\bf{k}})-(g_{_{V}}^2 -
4 g_{_{V'}}^2)U({\bf{k}})\bigg)\cos^2\theta_{_{W}}\nonumber\\
&\times&\int \frac{d^3{\bf q}}{(2\pi)^3}
\rho^0({\bf{x}},{\bf{q}})P_{2}(\cos\theta).
\end{eqnarray}
 The relations (\ref{I121})-(\ref{V121}) show that the CP occurs due to the forward scattering when the VDM is partially polarized and has a quadrapole distribution.
\section{APPENDIX B: CONTRIBUTION OF STOCKS PARAMETERS FROM DAMPING TERM}
The contribution of second term of (\ref{roij}) can be written as
\begin{eqnarray} \label{dampingterm1}
&&\frac{1}{2}\int_{-\infty}^{\infty}dt\,\left\langle [H_{I}^{0}(t),[H_{I}^{0}(0),D_{ij}({\bf{k}})]]\right\rangle =\int d{\bf{q_{1}}}\,d{\bf{q'_{1}}}\,d{\bf{p_{1}}}\,d{\bf{p'_{1}}}\,d{\bf{q_{2}}}\,d{\bf{q'_{2}}}\,d{\bf{p_{2}}}\,d{\bf{p'_{2}}}\nonumber\\
&\times&(2\pi)^7\,\delta^3({\bf{q'_{1}}}+{\bf{p'_{1}}}-{\bf{q_{1}}}-{\bf{p_{1}}})\delta^4(q'_{2}+p'_{2}-q_{2}-p_{2})\mathcal{M}(1)\mathcal{M}(2)\nonumber\\
&\times&\bigg\{(2\pi)^3p^{0}_{1}\,\delta_{is_{1}}\,\delta^3({\bf{p_{1}}}-{\bf{k}})\nonumber\\
&\times&
\left\langle b^{\dagger}_{r'_{2}}(q'_{2})b_{r_{2}}(q_{2})b^{\dagger}_{r'_{1}}(q'_{1})b_{r_{1}}(q_{1}) \right\rangle \left\langle a^{\dagger}_{s'_{2}}(p'_{2})a_{s_{2}}(p_{2})a^{\dagger}_{s'_{1}}(p'_{1})a_{j}(k) \right\rangle\nonumber\\
&&-(2\pi)^3p^{0}_{1}\,\delta_{is_{1}}\,\delta^3({\bf{p_{1}}}-{\bf{k}})\nonumber\\
&\times&\left\langle b^{\dagger}_{r'_{1}}(q'_{1})b_{r_{1}}(q_{1})b^{\dagger}_{r'_{2}}(q'_{2})b_{r_{2}}(q_{2}) \right\rangle \left\langle a^{\dagger}_{s'_{1}}(p'_{1})a_{j}(k)a^{\dagger}_{s'_{2}}(p'_{2})a_{s_{2}}(p_{2}) \right\rangle\nonumber\\
&&-(2\pi)^3p'^{0}_{1}\,\delta_{js'_{1}}\,\delta^3({\bf{p'_{1}}}-{\bf{k}})\nonumber\\
&\times&\left\langle b^{\dagger}_{r'_{2}}(q'_{2})b_{r_{1}}(q_{1})b^{\dagger}_{r'_{1}}(q'_{1})b_{r_{1}}(q_{1}) \right\rangle \left\langle a^{\dagger}_{s'_{2}}(p'_{2})a_{s_{2}}(p_{2})a^{\dagger}_{i}(k)a_{s_{1}}(p_{1}) \right\rangle\nonumber\\
&&+(2\pi)^3p'^{0}_{1}\,\delta_{js'_{1}}\,\delta^3({\bf{p'_{1}}}-{\bf{k}})\nonumber\\
&\times&\left\langle b^{\dagger}_{r'_{1}}(q'_{1})b_{r_{1}}(q_{1})b^{\dagger}_{r'_{2}}(q'_{2})b_{r_{2}}(q_{2}) \right\rangle\left\langle a^{\dagger}_{i}(k)a_{s_{1}}(p_{1})a^{\dagger}_{s'_{2}}(p'_{2})a_{s_{2}}(p_{2}) \right\rangle\bigg\},
\end{eqnarray}
where $\mathcal{M}=g^2_{_{V}}\cos^2 \theta_{_{W}}\mathcal{M}_{1}
+2g_{_{V}}g'_{_{V}}\cos^2 \theta_{_{W}}\mathcal{M}_{2}+g'^2_{_{V}}\cos^2 \theta_{_{W}}\mathcal{M}_{3}$. To calculate the above relation, we use the following expectation value for photon and VDM
\begin{eqnarray} \label{bbbb}
&&\left\langle a^{\dagger}_{s'_{1}}(p'_{1})a_{s_{1}}(p_{1})a^{\dagger}_{s'_{2}}(p'_{2})a_{s_{2}}(p_{2}) \right\rangle\nonumber\\
&=& 4p^{0}_{1}p^{0}_{2}(2\pi)^6\,\delta^3({\bf{p_{1}}}-{\bf{p'_{1}}})\,\delta^3({\bf{p_{2}}}-{\bf{p'_{2}}})\rho_{s_{1}s'_{1}}({\bf{p_{1}}})\rho_{s_{2}s'_{2}}({\bf{p_{2}}})\nonumber\\
&+& 4p^{0}_{1}p^{0}_{2}(2\pi)^6\,\delta^3({\bf{p_{1}}}-{\bf{p'_{2}}})\,\delta^3({\bf{p_{2}}}-{\bf{p'_{1}}})\rho_{s'_{1}s_{2}}({\bf{p_{2}}})[\delta_{s_{1}s'_{2}}+\rho_{s_{1}s'_{2}}({\bf{p_{1}}})],
\end{eqnarray}
and
\begin{eqnarray} \label{aaaa}
&&\left\langle b^{\dagger}_{r'_{1}}(q'_{1})b_{r_{1}}(q_{1})b^{\dagger}_{r'_{2}}(q'_{2})b_{r_{2}}(q_{2}) \right\rangle\nonumber\\
&=& 4q^{0}_{1}q^{0}_{2}(2\pi)^6\,\delta^3({\bf{q_{1}}}-{\bf{q'_{1}}})\,\delta^3({\bf{q_{2}}}-{\bf{q'_{2}}})\delta_{r_{1}r'_{1}}\delta_{r_{2}r'_{2}}\rho_{r_{1}r'_{1}}({\bf{q_{1}}})\rho_{r_{2}r'_{2}}({\bf{q_{2}}})\nonumber\\
&+& 4q^{0}_{1}q^{0}_{2}(2\pi)^6\,\delta^3({\bf{q_{1}}}-{\bf{q'_{2}}})\,\delta^3({\bf{q_{2}}}-{\bf{q'_{1}}})\delta_{r'_{1}r_{2}}\rho_{r'_{1}r_{2}}({\bf{q_{2}}})[\delta_{r_{1}r'_{2}}+\delta_{r_{1}r'_{2}}\rho_{r_{1}r'_{2}}({\bf{q_{1}}})],\nonumber\\
\end{eqnarray}
which leads to 
\begin{eqnarray} \label{dampingterm2}
&&\frac{1}{2}\int_{-\infty}^{\infty}dt\,\left\langle [H_{I}^{0}(t),[H_{I}^{0}(0),D_{ij}({\bf{k}})]]\right\rangle=\frac{1}{4}(2\pi)^3\delta^3(0)\int d{\bf{q}}\,d{\bf{q'}}\,d{\bf{p'}}\nonumber\\
&\times&
(2\pi)^4\,\delta^4(q'_{2}+p'_{2}-q_{2}-p_{2})
\mathcal{M}(q'r,p's'_{1},qr,ks_{1})\mathcal{M}(qr,ks'_{2},q'r,p's_{2})\nonumber\\
&\times&\bigg[\rho_{rr}({\bf{q}})\delta_{s_{2}s'_{1}}\bigg(\delta_{is_{1}}\rho_{s'_{2}j}({\bf{k}})+\delta_{js'_{2}}\rho_{is_{1}}({\bf{k}})\bigg)-2\rho_{rr}({\bf{q'}})\delta_{is_{1}}\delta_{js'_{2}}\rho_{s'_{1}s_{2}}({\bf{p'}})\bigg].\nonumber\\
\end{eqnarray}
From (\ref{roij}) and (\ref{dampingterm2}), the time evolution of the density matrix elements due to the damping term can be obtained as
\begin{eqnarray} \label{dampingterm2}
&&\frac{d}{dt}\rho_{ij}({\bf{x},{\bf{q}}})=-\frac{1}{4k^{0}}\int d{\bf{q}}\,d{\bf{p}}\nonumber\\
&\times&
\frac{1}{2E({\bf{q}}+{\bf{k}}-{\bf{p}})}(2\pi)\,\delta(E({\bf{q}}+{\bf{k}}-{\bf{p}})+p-E({\bf{q}})-k)\nonumber\\
&\times&\bigg[\rho_{rr}({\bf{x}},{\bf{q}})\delta_{s_{2}s'_{1}}\bigg(\delta_{is_{1}}\rho_{s'_{2}j}({{\bf{x}},\bf{k}})+\delta_{js'_{2}}\rho_{is_{1}}({{\bf{x}},\bf{k}})\bigg)\nonumber\\
&-&2\rho_{rr}({{\bf{x}},\bf{q'}})\delta_{is_{1}}\delta_{js'_{2}}\rho_{s'_{1}s_{2}}({{\bf{x}},\bf{p'}})\bigg] \mathcal{M}(q'r,p's'_{1},qr,ks_{1})\mathcal{M}(qr,ks'_{2},q'r,p's_{2}).\nonumber\\
\end{eqnarray}
Now by ignoring the recoil momentum of the VDM and using the following
approximations
\begin{eqnarray}\label{aprox}
	q'\approx q,
\end{eqnarray}
\begin{eqnarray}\label{aprox}
\rho_{rr}({\bf{x}},{\bf{q'}})\approx\rho_{rr}({\bf{x}},{\bf{q}}),
\end{eqnarray}
\begin{eqnarray}\label{aprox}
E({\bf{q}}+{\bf{k}}-{\bf{p}})\simeq E({\bf{q}}),
\end{eqnarray}
\begin{eqnarray}\label{aprox}
\delta(E({\bf{q}}+{\bf{k}}-{\bf{p}})+p-E({\bf{q}})-k)\simeq\delta(p-k),
\end{eqnarray}
one can cast the  Stockes parameters into
\begin{eqnarray} \label{Idot}
&&\dot{I}({\bf{k}})=\dot{\rho}_{11}({\bf{k}})+\dot{\rho}_{22}({\bf{k}})=-\frac{\pi}{4k^{0} m_{_{V}}}\int d{\bf{q}}\,d{\bf{p}}\,\delta(p-k)\,\sum_{r=1}^{3}\rho_{rr}({\bf{x}},{\bf{q}})\,\bigg\{I({\bf{k}})\nonumber\\
&\times&\bigg[\mathcal{M}(qr,p'(1),qr,k(1))\mathcal{M}(qr,k(1),qr,p(1))+\mathcal{M}(qr,p'(2),qr,k(1))\mathcal{M}(qr,k(1),qr,p(2))\nonumber\\
&+&\mathcal{M}(qr,p'(1),qr,k(2))\mathcal{M}(qr,k(2),qr,p(1))+\mathcal{M}(qr,p'(2),qr,k(2))\mathcal{M}(qr,k(2),qr,p(2))\bigg]\nonumber\\
&+&Q({\bf{k}})\nonumber\\
&\times&\bigg[\mathcal{M}(qr,p'(1),qr,k(1))\mathcal{M}(qr,k(1),qr,p(1))+\mathcal{M}(qr,p'(2),qr,k(1))\mathcal{M}(qr,k(1),qr,p(2))\nonumber\\
&-&\mathcal{M}(qr,p'(1),qr,k(2))\mathcal{M}(qr,k(2),qr,p(1))-\mathcal{M}(qr,p'(2),qr,k(2))\mathcal{M}(qr,k(2),qr,p(2))\bigg]\nonumber\\
&+&U({\bf{k}})\nonumber\\
&\times&\bigg[\mathcal{M}(qr,p'(1),qr,k(2))\mathcal{M}(qr,k(1),qr,p(1))+\mathcal{M}(qr,p'(2),qr,k(2))\mathcal{M}(qr,k(1),qr,p(2))\nonumber\\
&+&\mathcal{M}(qr,p'(1),qr,k(1))\mathcal{M}(qr,k(2),qr,p(1))+\mathcal{M}(qr,p'(2),qr,k(1))\mathcal{M}(qr,k(2),qr,p(2))\bigg]\nonumber\\
&+& iV({\bf{k}})\nonumber\\
&\times&\bigg[\mathcal{M}(qr,p'(1),qr,k(1))\mathcal{M}(qr,k(2),qr,p(1))+\mathcal{M}(qr,p'(2),qr,k(1))\mathcal{M}(qr,k(2),qr,p(2))\nonumber\\
&-&\mathcal{M}(qr,p'(1),qr,k(2))\mathcal{M}(qr,k(1),qr,p(1))-\mathcal{M}(qr,p'(2),qr,k(2))\mathcal{M}(qr,k(1),qr,p(2))\bigg] \nonumber\\
&-&I({\bf{p}})\nonumber\\
&\times&\bigg[\mathcal{M}(qr,p'(1),qr,k(1))\mathcal{M}(qr,k(1),qr,p(1))+\mathcal{M}(qr,p'(1),qr,k(2))\mathcal{M}(qr,k(2),qr,p(1))\nonumber\\
&+&\mathcal{M}(qr,p'(2),qr,k(1))\mathcal{M}(qr,k(1),qr,p(2))+\mathcal{M}(qr,p'(2),qr,k(2))\mathcal{M}(qr,k(2),qr,p(2))\bigg]\nonumber\\
&-&Q({\bf{p}})\nonumber\\
&\times&\bigg[\mathcal{M}(qr,p'(1),qr,k(1))\mathcal{M}(qr,k(1),qr,p(1))+\mathcal{M}(qr,p'(1),qr,k(2))\mathcal{M}(qr,k(2),qr,p(1))\nonumber\\
&-&\mathcal{M}(qr,p'(2),qr,k(1))\mathcal{M}(qr,k(1),qr,p(2))-\mathcal{M}(qr,p'(2),qr,k(2))\mathcal{M}(qr,k(2),qr,p(2))\bigg]\nonumber\\
&-&U({\bf{p}})\nonumber\\
&\times&\bigg[\mathcal{M}(qr,p'(1),qr,k(1))\mathcal{M}(qr,k(1),qr,p(2))+\mathcal{M}(qr,p'(1),qr,k(2))\mathcal{M}(qr,k(2),qr,p(2))\nonumber\\
&+&\mathcal{M}(qr,p'(2),qr,k(1))\mathcal{M}(qr,k(1),qr,p(1))+\mathcal{M}(qr,p'(2),qr,k(2))\mathcal{M}(qr,k(2),qr,p(1))\bigg]\nonumber\\
&-& iV({\bf{p}})\nonumber\\
&\times&\bigg[\mathcal{M}(qr,p'(2),qr,k(1))\mathcal{M}(qr,k(1),qr,p(1))+\mathcal{M}(qr,p'(2),qr,k(2))\mathcal{M}(qr,k(2),qr,p(1))\nonumber\\
&-&\mathcal{M}(qr,p'(1),qr,k(1))\mathcal{M}(qr,k(1),qr,p(2))-\mathcal{M}(qr,p'(1),qr,k(2))\mathcal{M}(qr,k(2),qr,p(2))\bigg]\bigg\},\nonumber\\
\end{eqnarray}

\begin{eqnarray} \label{Qdot}
&&\dot{Q}({\bf{k}})=\dot{\rho}_{11}({\bf{k}})-\dot{\rho}_{22}({\bf{k}})=-\frac{\pi}{4k^{0} m_{_{V}}}\int d{\bf{q}}\,d{\bf{p}}\,\delta(p-k)\,\sum_{r=1}^{3}\rho_{rr}({\bf{x}},{\bf{q}}) \bigg\{I({\bf{k}})\nonumber\\
&\times&\bigg[\mathcal{M}(qr,p'(1),qr,k(1))\mathcal{M}(qr,k(1),qr,p(1))+\mathcal{M}(qr,p'(2),qr,k(1))\mathcal{M}(qr,k(1),qr,p(2))\nonumber\\
&-&\mathcal{M}(qr,p'(1),qr,k(2))\mathcal{M}(qr,k(2),qr,p(1))-\mathcal{M}(qr,p'(2),qr,k(2))\mathcal{M}(qr,k(2),qr,p(2))\bigg]\nonumber\\
&+&Q({\bf{k}})\nonumber\\
&\times&\bigg[\mathcal{M}(qr,p'(1),qr,k(1))\mathcal{M}(qr,k(1),qr,p(1))+\mathcal{M}(qr,p'(2),qr,k(1))\mathcal{M}(qr,k(1),qr,p(2))\nonumber\\
&+&\mathcal{M}(qr,p'(1),qr,k(2))\mathcal{M}(qr,k(2),qr,p(1))+\mathcal{M}(qr,p'(2),qr,k(2))\mathcal{M}(qr,k(2),qr,p(2))\bigg]\nonumber\\
&-&I({\bf{p}})\nonumber\\
&\times&\bigg[\mathcal{M}(qr,p'(1),qr,k(1))\mathcal{M}(qr,k(1),qr,p(1))-\mathcal{M}(qr,p'(1),qr,k(2))\mathcal{M}(qr,k(2),qr,p(1))\nonumber\\
&+&\mathcal{M}(qr,p'(2),qr,k(1))\mathcal{M}(qr,k(1),qr,p(2))-\mathcal{M}(qr,p'(2),qr,k(2))\mathcal{M}(qr,k(2),qr,p(2))\bigg]\nonumber\\
&-&Q({\bf{p}})\nonumber\\
&\times&\bigg[\mathcal{M}(qr,p'(1),qr,k(1))\mathcal{M}(qr,k(1),qr,p(1))-\mathcal{M}(qr,p'(1),qr,k(2))\mathcal{M}(qr,k(2),qr,p(1))\nonumber\\
&-&\mathcal{M}(qr,p'(2),qr,k(1))\mathcal{M}(qr,k(1),qr,p(2))+\mathcal{M}(qr,p'(2),qr,k(2))\mathcal{M}(qr,k(2),qr,p(2))\bigg]\nonumber\\
&-&U({\bf{p}})\nonumber\\
&\times&\bigg[\mathcal{M}(qr,p'(2),qr,k(1))\mathcal{M}(qr,k(1),qr,p(1))-\mathcal{M}(qr,p'(2),qr,k(2))\mathcal{M}(qr,k(2),qr,p(1))\nonumber\\
&+&\mathcal{M}(qr,p'(1),qr,k(1))\mathcal{M}(qr,k(1),qr,p(2))-\mathcal{M}(qr,p'(1),qr,k(2))\mathcal{M}(qr,k(2),qr,p(2))\bigg]\nonumber\\
&-&iV({\bf{p}})\nonumber\\
&\times&\bigg[\mathcal{M}(qr,p'(2),qr,k(1))\mathcal{M}(qr,k(1),qr,p(1))-\mathcal{M}(qr,p'(2),qr,k(2))\mathcal{M}(qr,k(2),qr,p(1))\nonumber\\
&-&\mathcal{M}(qr,p'(1),qr,k(1))\mathcal{M}(qr,k(1),qr,p(2))
+\mathcal{M}(qr,p'(1),qr,k(2))\mathcal{M}(qr,k(2),qr,p(2))\bigg]\bigg\},\nonumber\\
\end{eqnarray}

\begin{eqnarray} \label{Udot}
&&\dot{U}({\bf{k}})=\dot{\rho}_{12}({\bf{k}})+\dot{\rho}_{21}({\bf{k}})=-\frac{\pi}{4k^{0} m_{_{V}}}\int d{\bf{q}}\,d{\bf{p}}\,\delta(p-k)\,\sum_{r=1}^{3}\rho_{rr}({\bf{x}},{\bf{q}})\bigg\{I({\bf{k}})\nonumber\\
&\times&\bigg[\mathcal{M}(qr,p'(1),qr,k(2))\mathcal{M}(qr,k(1),qr,p(1))+\mathcal{M}(qr,p'(2),qr,k(2))\mathcal{M}(qr,k(1),qr,p(2))\nonumber\\
&+&\mathcal{M}(qr,p'(1),qr,k(1))\mathcal{M}(qr,k(2),qr,p(1))+\mathcal{M}(qr,p'(2),qr,k(1))\mathcal{M}(qr,k(2),qr,p(2))\bigg]\nonumber\\
&+&U({\bf{k}})\nonumber\\
&\times&\bigg[\mathcal{M}(qr,p'(1),qr,k(2))\mathcal{M}(qr,k(2),qr,p(1))+\mathcal{M}(qr,p'(2),qr,k(2))\mathcal{M}(qr,k(2),qr,p(2))\nonumber\\
&+&\mathcal{M}(qr,p'(2),qr,k(1))\mathcal{M}(qr,k(1),qr,p(2))+\mathcal{M}(qr,p'(1),qr,k(1))\mathcal{M}(qr,k(1),qr,p(1))\bigg]\nonumber\\
&-&I({\bf{p}})\nonumber\\
&\times&\bigg[\mathcal{M}(qr,p'(1),qr,k(2))\mathcal{M}(qr,k(1),qr,p(1))+\mathcal{M}(qr,p'(1),qr,k(1))\mathcal{M}(qr,k(2),qr,p(1))\nonumber\\
&+&\mathcal{M}(qr,p'(2),qr,k(2))\mathcal{M}(qr,k(1),qr,p(2))+\mathcal{M}(qr,p'(2),qr,k(1))\mathcal{M}(qr,k(2),qr,p(2))\bigg]\nonumber\\
&-&Q({\bf{p}})\nonumber\\
&\times&\bigg[\mathcal{M}(qr,p'(1),qr,k(2))\mathcal{M}(qr,k(1),qr,p(1))+\mathcal{M}(qr,p'(1),qr,k(1))\mathcal{M}(qr,k(2),qr,p(1))\nonumber\\
&-&\mathcal{M}(qr,p'(2),qr,k(2))\mathcal{M}(qr,k(1),qr,p(2))-\mathcal{M}(qr,p'(2),qr,k(1))\mathcal{M}(qr,k(2),qr,p(2))\bigg]\nonumber\\
&-&U({\bf{p}})\nonumber\\
&\times&\bigg[\mathcal{M}(qr,p'(1),qr,k(1))\mathcal{M}(qr,k(2),qr,p(2))+\mathcal{M}(qr,p'(1),qr,k(2))\mathcal{M}(qr,k(1),qr,p(2))\nonumber\\
&+&\mathcal{M}(qr,p'(2),qr,k(2))\mathcal{M}(qr,k(1),qr,p(1))+\mathcal{M}(qr,p'(2),qr,k(1))\mathcal{M}(qr,k(2),qr,p(1))\bigg]\nonumber\\
&-&iV({\bf{p}})\nonumber\\
&\times&\bigg[\mathcal{M}(qr,p'(2),qr,k(2))\mathcal{M}(qr,k(1),qr,p(1))+\mathcal{M}(qr,p'(2),qr,k(1))\mathcal{M}(qr,k(2),qr,p(1))\nonumber\\
&-&\mathcal{M}(qr,p'(1),qr,k(2))\mathcal{M}(qr,k(1),qr,p(2))-\mathcal{M}(qr,p'(1),qr,k(1))\mathcal{M}(qr,k(2),qr,p(2))\bigg]\bigg\},\nonumber\\
\end{eqnarray}
and
\begin{eqnarray} \label{Vdot}
&&\dot{V}({\bf{k}})=i(\dot{\rho}_{12}({\bf{k}})-\dot{\rho}_{21}({\bf{k}}))=-\frac{i\pi}{4k^{0} m_{_{V}}}\int d{\bf{q}}\,d{\bf{p}}\,\delta(p-k)\,\sum_{r=1}^{3}\rho_{rr}({\bf{x}},{\bf{q}}) \bigg\{I({\bf{k}})\nonumber\\
&\times&\bigg[\mathcal{M}(qr,p'(1),qr,k(1))\mathcal{M}(qr,k(2),qr,p(1))+\mathcal{M}(qr,p'(2),qr,k(1))\mathcal{M}(qr,k(2),qr,p(2))\nonumber\\
&-&\mathcal{M}(qr,p'(1),qr,k(2))\mathcal{M}(qr,k(1),qr,p(1))-\mathcal{M}(qr,p'(2),qr,k(2))\mathcal{M}(qr,k(1),qr,p(2))\bigg]\nonumber\\
&+&iV({\bf{k}})\nonumber\\
&\times&\bigg[\mathcal{M}(qr,p'(1),qr,k(1))\mathcal{M}(qr,k(1),qr,p(1))+\mathcal{M}(qr,p'(2),qr,k(1))\mathcal{M}(qr,k(1),qr,p(2))\nonumber\\
&+&\mathcal{M}(qr,p'(1),qr,k(2))\mathcal{M}(qr,k(2),qr,p(1))+\mathcal{M}(qr,p'(2),qr,k(2))\mathcal{M}(qr,k(2),qr,p(2))\bigg]\nonumber\\
&-&I({\bf{p}})\nonumber\\
&\times&\bigg[\mathcal{M}(qr,p'(1),qr,k(1))\mathcal{M}(qr,k(2),qr,p(1))-\mathcal{M}(qr,p'(1),qr,k(2))\mathcal{M}(qr,k(1),qr,p(1))\nonumber\\
&+&\mathcal{M}(qr,p'(2),qr,k(1))\mathcal{M}(qr,k(2),qr,p(2))-\mathcal{M}(qr,p'(2),qr,k(2))\mathcal{M}(qr,k(1),qr,p(2))\bigg]\nonumber\\
&-&Q({\bf{p}})\nonumber\\
&\times&\bigg[\mathcal{M}(qr,p'(1),qr,k(1))\mathcal{M}(qr,k(2),qr,p(1))-\mathcal{M}(qr,p'(1),qr,k(2))\mathcal{M}(qr,k(1),qr,p(1))\nonumber\\
&-&\mathcal{M}(qr,p'(2),qr,k(1))\mathcal{M}(qr,k(2),qr,p(2))+\mathcal{M}(qr,p'(2),qr,k(2))\mathcal{M}(qr,k(1),qr,p(2))\bigg]\nonumber\\
&-&U({\bf{p}})\nonumber\\
&\times&\bigg[\mathcal{M}(qr,p'(1),qr,k(1))\mathcal{M}(qr,k(2),qr,p(2))-\mathcal{M}(qr,p'(1),qr,k(2))\mathcal{M}(qr,k(1),qr,p(2))\nonumber\\
&+&\mathcal{M}(qr,p'(2),qr,k(1))\mathcal{M}(qr,k(2),qr,p(1))-\mathcal{M}(qr,p'(2),qr,k(2))\mathcal{M}(qr,k(1),qr,p(1))\bigg]\nonumber\\
&-&iV({\bf{p}})\nonumber\\
&\times&\bigg[\mathcal{M}(qr,p'(2),qr,k(1))\mathcal{M}(qr,k(2),qr,p(1))-\mathcal{M}(qr,p'(2),qr,k(2))\mathcal{M}(qr,k(1),qr,p(1))\nonumber\\
&-&\mathcal{M}(qr,p'(1),qr,k(1))\mathcal{M}(qr,k(2),qr,p(2))
+\mathcal{M}(qr,p'(1),qr,k(2))\mathcal{M}(qr,k(1),qr,p(2))\bigg]\bigg\}.\nonumber\\
\end{eqnarray}
If we consider only the first term of the propagator and Using the relations (\ref{Eq1})-(\ref{Eq2}) for the VDM and the incoming photon and
\begin{eqnarray}\label{Eq3}
p&=&(p^0,| {\bf{p}}|\sin\alpha'\cos\beta',| {\bf{p}}|\sin\alpha'\sin\beta',| {\bf{p}}|\cos\alpha'),\\
\varepsilon_{1}(p)&=&(0, \cos\alpha'\cos\beta',\cos\alpha'\sin\beta',-\sin\alpha'),\\
\varepsilon_{2}(p)&=&(0, -\sin\beta',\cos\beta',0),
\end{eqnarray}
for the outgoing photon, then by applying the non-relativistic limits $q^0=m_{_{V}}$ and $(m^2_{_{V}}-m^2_{_{V'}})^2\gg 4m^2_{_{V}}k^{0^2}$, the Stockes parameters can be written as
\begin{eqnarray} \label{stoke11}
\dot{I}({\bf{k}})=-\frac{16}{3}\frac{n_{DM}(x)}{m_{_{V}}^{2}}\frac{k^{0^{4}}\cos^4\theta_{_{W}}}{(m_{_{V'}}^{2}-m_{_{V}}^{2})^2}(g_{_{V}}^{2}+4g_{_{V}}^{'2})^2\bigg(I({\bf{k}})-\int \frac{d\Omega'}{4\pi}I({\bf{p}})\bigg)
\end{eqnarray}
\begin{eqnarray} \label{stoke10}
\dot{Q}({\bf{k}})&=&-\frac{4}{15}\frac{n_{DM}(x)}{m_{_{V}}^{2}}\frac{k^{0^{4}}\sin^2\alpha\cos^4\theta_{_{W}}}{(m_{_{V'}}^{2}-m_{_{V}}^{2})^2}(g_{_{V}}^{4}-16g_{_{V}}^{'4})\nonumber\\ &\times &\,\bigg(I({\bf{k}})\int\frac{d\Omega'}{4\pi}P_{2}(\cos\alpha')-\int \frac{d\Omega'}{4\pi}P_{2}(\cos\alpha')I({\bf{p}})\bigg),
\end{eqnarray}
\begin{eqnarray} \label{stoke10}
\dot{U}({\bf{k}})&=&-\frac{16}{15}\frac{n_{DM}(x)}{m_{_{V}}^{2}}\frac{k^{0^{4}}\sin^2\alpha\cos^4\theta_{_{W}}}{(m_{_{V'}}^{2}-m_{_{V}}^{2})^2}g_{_{V}}g'_{_{V}}(g_{_{V}}^{2}+4g_{_{V}}^{'2})\nonumber\\ &\times &\,\bigg(I({\bf{k}})\int\frac{d\Omega'}{4\pi}P_{2}(\cos\alpha')-\int \frac{d\Omega'}{4\pi}P_{2}(\cos\alpha')I({\bf{p}})\bigg),
\end{eqnarray}
and
\begin{eqnarray} \label{stoke12}
\dot{V}({\bf{k}})=0.
\end{eqnarray}
Now according to the orthonormality condition for the Legendre polynomials one has
\begin{eqnarray} \label{stoke11}
\dot{I}({\bf{k}})=-\frac{16}{3}\frac{n_{DM}(x)}{m_{_{V}}^{2}}\frac{k^{0^{4}}\cos^4\theta_{_{W}}}{(m_{_{V'}}^{2}-m_{_{V}}^{2})^2}(g_{_{V}}^{2}+4g_{_{V}}^{'2})^2\,I({\bf{k}}),
\end{eqnarray}
\begin{eqnarray} \label{stoke10}
\dot{Q}({\bf{k}})=-\frac{4}{15}\frac{n_{DM}(x)}{m_{_{V}}^{2}}\frac{k^{0^{4}}\sin^2\alpha\cos^4\theta_{_{W}}}{(m_{_{V'}}^{2}-m_{_{V}}^{2})^2}(g_{_{V}}^{4}-16g_{_{V}}^{'4})\,I_{2}({\bf{p}}),
\end{eqnarray}
\begin{eqnarray} \label{stoke10}
\dot{U}({\bf{k}})=-\frac{16}{15}\frac{n_{DM}(x)}{m_{_{V}}^{2}}\frac{k^{0^{4}}\sin^2\alpha\cos^4\theta_{_{W}}}{(m_{_{V'}}^{2}-m_{_{V}}^{2})^2}g_{_{V}}g'_{_{V}}(g_{_{V}}^{2}+4g_{_{V}}^{'2})\,I_{2}({\bf{p}}),
\end{eqnarray}
and
\begin{eqnarray} \label{stoke12}
\dot{V}({\bf{k}})=0,
\end{eqnarray}
where
\begin{eqnarray} \label{stoke10}
I_{2}({\bf{p}})=\int \frac{d\Omega'}{4\pi}P_{2}(\cos\alpha')I({\bf{p}}).
\end{eqnarray}
The above relations show that the largest order in the damping term produces only a linear polarization. 
One can compare contribution of the VDM-photon scattering with the Compton scattering as
\begin{equation}\label{Op4}
\frac{\dot{\tau}_{DM}}{\dot{\tau}_{e\gamma}}\approx\frac{n_{DM}}{n_e}\frac{\langle\sigma v_{rel}\rangle_{ann}}{\sigma_T} \left( \frac{k^0}{m_{_{V}}}\right) ^4,
\end{equation}
which for $\left( \frac{k^0}{m_{_{V}}}\right) ^4 \ll 1$  is too small to be considered in comparison with the Compton scattering. 
\end{document}